\documentclass[pdflatex,sn-aps]{sn-jnl}

\PassOptionsToPackage{table}{xcolor}
\usepackage{xcolor}    

\usepackage{graphicx}%
\usepackage{multirow}%
\usepackage{amsmath,amssymb,amsfonts}%
\usepackage{amsthm}%
\usepackage{mathrsfs}%
\usepackage[title]{appendix}%
\usepackage{textcomp}%
\usepackage{manyfoot}%
\usepackage{booktabs}%
\usepackage{algorithm}%
\usepackage{algorithmicx}%
\usepackage{algpseudocode}%
\usepackage{listings}%
\usepackage{tcolorbox}

\usepackage{orcidlink}
\usepackage[framemethod=TikZ]{mdframed}
\DeclareMathOperator*{\argmax}{arg\,max}
\usepackage{braket}

\usepackage{lscape} 
\usepackage{longtable}
\usepackage{tabularx}
\usepackage{pdflscape}
\usepackage{caption}      

\usepackage{tikz}
\usetikzlibrary{shapes.geometric, arrows.meta, positioning}
\tikzset{
  layer/.style={
    rectangle,
    draw=black,
    fill=gray!10,
    minimum width=\textwidth,
    minimum height=1cm,
    align=center,
    font=\small,
    inner sep=3pt
  }
}

\usepackage{enumitem}

\theoremstyle{thmstyleone}

\theoremstyle{thmstyletwo}

\theoremstyle{thmstylethree}

\begin{document}

\title[Quantum Computer Benchmarking: An Explorative Systematic Literature Review]{Quantum Computer Benchmarking: An Explorative Systematic Literature Review}

\author*[1]{\fnm{Tobias} \sur{Rohe}\orcidlink{0009-0003-3283-0586}}\email{tobias.rohe@ifi.lmu.de}
\author[1, 2]{\fnm{Federico} \sur{Harjes Ruiloba}}
\author[1]{\fnm{Sabrina} \sur{Egger}}
\author[3]{\fnm{Sebastian} \sur{von Beck}}
\author[1,4]{\fnm{Jonas} \sur{Stein}\orcidlink{0000-0001-5727-9151}}
\author[1]{\fnm{Claudia} \sur{Linnhoff-Popien}\orcidlink{0000-0001-6284-9286}}

\affil[1]{\orgdiv{Institute for Computer Science}, \orgname{LMU Munich}, \orgaddress{\street{Oettingenstraße 67}, \city{Munich}, \postcode{80538}, \state{Bavaria}, \country{Germany}}}

\affil[2]{\orgdiv{relAI – Konrad Zuse School of Excellence in Reliable AI}, \orgname{LMU Munich} \orgaddress{\street{Akademiestraße 7}, \city{Munich}, \postcode{80799}, \state{Bavaria}, \country{Germany}}}

\affil[3]{\orgdiv{Institute for Human Capital Management}, \orgname{LMU Munich}, \orgaddress{\street{Schackstraße 4}, \city{Munich}, \postcode{80539}, \state{Bavaria}, \country{Germany}}}

\affil[4]{\orgname{Aqarios GmbH}, \orgaddress{\street{Prinzregentenstraße 120}, \city{Munich}, \postcode{81677}, \state{Bavaria}, \country{Germany}}}

\abstract{As quantum computing (QC) continues to evolve in hardware and software, measuring progress in this complex and diverse field remains a challenge. To track progress, uncover bottlenecks, and evaluate community efforts, benchmarks play a crucial role. But which benchmarking approach best addresses the diverse perspectives of QC stakeholders? We conducted the most comprehensive systematic literature review of this area to date, combining NLP-based clustering with expert analysis to develop a novel taxonomy and definitions for QC benchmarking, aligned with the quantum stack and its stakeholders. In addition to organizing benchmarks in distinct hardware, software, and application focused categories, our taxonomy hierarchically classifies benchmarking protocols in clearly defined subcategories. We develop standard terminology and map the interdependencies of benchmark categories to create a holistic, unified picture of the quantum benchmarking landscape. Our analysis reveals recurring design patterns, exposes research gaps, and clarifies how benchmarking methods serve different stakeholders. By structuring the field and providing a common language, our work offers a foundation for  coherent benchmark development, fairer evaluation, and stronger cross-disciplinary collaboration in QC.}


\maketitle

\section{Introduction}\label{introduction}
Quantum computers can process and store information within a high-dimensional state space~\cite{preskill2023quantum}. Combined with promising algorithms with up to exponential speed-ups, such as Shor's~\cite{shor1999polynomial}, quantum computing (QC) is widely regarded as a disruptive technology. However, recently, expectations and opinions regarding QC’s progress and readiness have diverged~\cite{schuld2022quantum, hoefler2023disentangling}. While it is possible to superficially assess QC’s progress by looking only at the number of qubits, as marketing departments and media often do, this perspective is misleading and incomplete. It’s not just about the size and scaling of QC but also about many more intricate factors like the quality of qubits, their gates, and their connectivity~\cite{preskill2018quantum}. A diverse set of complex benchmarking techniques is required to holistically assess progress, identify bottlenecks, and evaluate QC’s future potential~\cite{lubinski2023application, murphy2019controlling}. This implies that different aspects and perspectives require distinct benchmarking approaches~\cite{proctor2025benchmarking}. No single benchmark can comprehensively address all aspects of QC assessment. At the same time, QC benchmarking is challenging due to variations in hardware platforms, target metrics, and performance dependencies on specific applications and algorithms~\cite{resch2021benchmarking}. Moreover, the hybrid nature of today’s quantum applications, which typically involve both classical and quantum hardware, further complicates the design of fair benchmarks. Literature currently lacks clear definitions and standardized benchmarking methodologies~\cite{bowles2024better}.

This review addresses the pressing need for a unified, structured framework that categorizes and contextualizes gate-based QC benchmarks across the quantum stack, without neglecting the various stakeholders involved. We conducted a systematic literature search using a Boolean keyword-based strategy across multiple academic databases. Using inclusion and exclusion criteria, we conducted a double-blind review of $401$ titles and abstracts, identifying $175$ papers for full-text assessment. Following this stage, an additional $34$ papers were excluded, leading to a final selection of $141$ studies for initial analysis, which we further extended via forward-backward search. For clustering we combined natural language processing (NLP) using BERTopic~\cite{grootendorst2022bertopic} with rigorous manual refinement to produce a novel taxonomy. This taxonomy is paired with standardized terminology and definitions to address the fragmented language in the field and with an analysis of interdependencies and recurring design patterns across benchmark types. Beyond structuring the field, our work identifies research gaps and underexplored areas, providing clear directions for future benchmarking research. The result is an up-to-date, stakeholder-oriented reference that supports the understanding, design, comparison, and fair evaluation of QC systems.

The remainder of this paper is structured as follows: Sec.~\ref{background} provides background information, introducing key concepts in benchmarking (Sec.~\ref{benchmarks}) and outlining the QC stack (Sec.~\ref{qc_stack}). Sec.~\ref{relatedwork} reviews related work, while Sec.~\ref{methodology} describes the methodology used in this study, detailing the systematic literature review process, clustering techniques, and data synthesis. Sec.~\ref{results} presents the results, structured into three perspectives: Hardware Focus (Sec.~\ref{hardware}), Software Focus (Sec.~\ref{software}), and Application Focus (Sec.~\ref{application}), highlighting key benchmarking approaches in each domain. Sec.~\ref{interconnectedness} discusses the interconnectedness of benchmarking methods and their relationship to each other. Finally, Sec.~\ref{conclusion} concludes the paper, summarizing key insights and outlining directions for future research.

\section{Background}\label{background}
In the following, we first define QC benchmarks and outline their key characteristics and objectives, providing a foundation for subsequent discussions. We then introduce the layered structure of the QC stack, which serves as a conceptual framework for locating and contextualizing different benchmarking approaches within the broader system architecture.

\subsection{Benchmarks}\label{benchmarks}
Similar to other rapidly advancing scientific domains, the field of QC benchmarking lacks well-established definitions and terminology. Acuaviva et al. (2024) recently introduced several definitions of key terms in QC benchmarking, and we adopt their definition of QC benchmarks as it is both concise and comprehensive:

\begin{quote}
\textit{“A test, or set of tests, that aims to measure or evaluate the performance, efficiency, or other properties of a quantum processor or hardware component for a certain task.”}
\begin{flushright}
(Acuaviva et al. (2024)~\cite{acuaviva2024benchmarking}, p.~10)
\end{flushright}
\end{quote}

This definition emphasizes the purpose of quantifying device performance and its components, and forms the basis for comparisons between different devices and tracking their development over time. The definition is sufficiently broad, allowing it to encompass a wide range of benchmarks. As observed in our research, we aim to clearly distinguish QC benchmarking from quantum (computing) verification, which focuses on determining whether a quantum device functions correctly—an important distinction that is often overlooked.

To evaluate benchmarks and assess their robustness, several key characteristics have been identified in literature~\cite{v2015build, dai2019benchmarking, donkers2022qpack, tomesh2022supermarq, acuaviva2024benchmarking}. Drawing from these, we have synthesized common principles that capture the most widely agreed-upon characteristics. While the relative importance of these principles depends on the benchmark’s specific purpose, additional principles may also be relevant.

\begin{enumerate}[label=(\alph*)]
    \item \textbf{Relevance:} Benchmarks must be relevant and meaningful, designed specifically to address their intended purpose. This purpose may involve evaluating aspects such as speed, scalability, accuracy, energy consumption, or quality, depending on the benchmark's specific goals. Benchmarks should be representative, capturing diverse program structures and resource requirements to provide comprehensive insights into system performance under different conditions. Metrics should target the critical features of interest and be widely accepted within the scientific and industrial communities.
    \item \textbf{Fairness:} Benchmarks should facilitate fair cross-system comparisons, ensuring that different test configurations are evaluated solely on their intrinsic performance, free from biases or artificial constraints. Fairness requires consistency in benchmarking conditions while supporting diverse configurations, which inherently involves trade-offs. Establishing a common ground is essential for equitable and transparent evaluation.
    \item \textbf{Reproducibility:} Benchmarks must produce results that are consistent when run with the same test configuration, enabling replication and validation. Reproducibility is challenging in practice due to the difficulty of maintaining steady-state conditions and replicating dynamic real-world usage scenarios. It requires benchmarks to rank systems consistently based on the evaluated property, ensuring that better-performing systems achieve superior results across multiple runs. Consistency in benchmarks demands the use of uniform units and precise definitions across different systems and configurations. A lack of consistency undermines the verifiability and relevance of benchmarks.
    \item \textbf{Usability:} Usability is a critical attribute, as user-friendly benchmarks are more likely to be adopted and reduce the likelihood of execution errors. Additionally, benchmarks should be cost-effective to lower adoption barriers. Together, usability and affordability enhance accessibility, promoting widespread and sustained benchmark usage for long-term performance assessment.
    \item \textbf{Scalability:} Benchmarks should accommodate a wide range of system sizes, reflecting the progression of QC from small-scale noisy intermediate-scale quantum (NISQ)~\cite{preskill2018quantum} devices to large-scale fault-tolerant quantum computers. This naturally depends on the benchmark’s focus, as certain approaches, such as tomography-based benchmarks, do in general not scale well. Scalable benchmarks should use parameterizable size settings and ensure that performance metrics remain efficiently verifiable, avoiding dependence on classical simulations that scale exponentially with the number of qubits.
    \item \textbf{Transparency:} Benchmark metrics should be transparent and understandable, instilling confidence in their accuracy. They must be verifiable through well-established methodologies to ensure reliability. In academia, benchmarks undergo peer reviews to validate results, while in industry, self-validation mechanisms are crucial to support independent corroboration and usability. Additionally, linear metrics are valuable due to their intuitive nature, where proportional increases in metric values reflect proportional improvements in performance. 
\end{enumerate}

Regardless of these properties, deviations from expected benchmark performance can still arise during application execution. Importantly, such deviations do not necessarily reflect a lack of benchmark relevance per se. Rather, they often result from a mismatch between the benchmark’s intended evaluation goal and the actual application context, underscoring the challenge of correctly selecting and applying even well-designed benchmarks.

Furthermore, while scalability in terms of system size is an essential benchmark attribute, benchmarks must also adapt to the broader evolution of QC technology. As quantum hardware progresses from small-scale NISQ devices to large-scale fault-tolerant architectures, benchmarks must evolve accordingly to remain valid~\cite{acuaviva2024benchmarking}. This need for adaptability goes beyond simply scaling the number of qubits and includes improvements in gate fidelity, the introduction of new quantum operations, and shifts in application paradigms, thus posing a significant ongoing challenge for benchmark development.

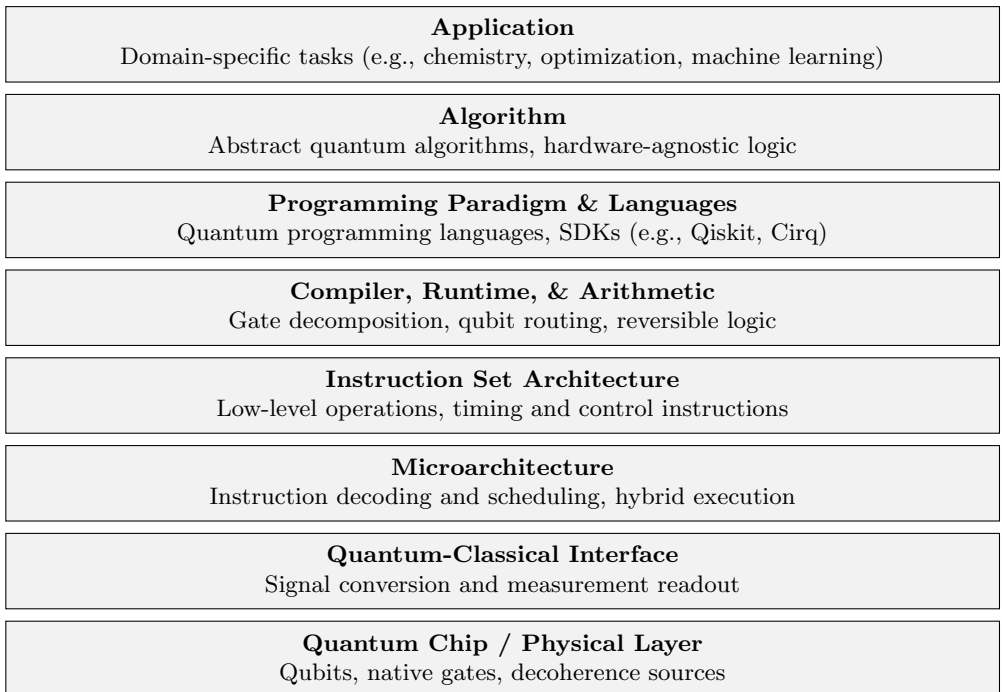
\begin{figure*}[t]
\centering
\begin{tikzpicture}[
  node distance=0.15cm
]

\node[layer] (application) {\textbf{Application} \\ Domain-specific tasks (e.g., chemistry, optimization, machine learning)};
\node[layer, below=of application] (algorithm) {\textbf{Algorithm} \\ Abstract quantum algorithms, hardware-agnostic logic};
\node[layer, below=of algorithm] (language) {\textbf{Programming Paradigm \& Languages} \\ Quantum programming languages, SDKs (e.g., Qiskit, Cirq)};
\node[layer, below=of language] (compiler) {\textbf{Compiler, Runtime, \& Arithmetic} \\ Gate decomposition, qubit routing, reversible logic};
\node[layer, below=of compiler] (isa) {\textbf{Instruction Set Architecture} \\ Low-level operations, timing and control instructions};
\node[layer, below=of isa] (micro) {\textbf{Microarchitecture} \\ Instruction decoding and scheduling, hybrid execution};
\node[layer, below=of micro] (interface) {\textbf{Quantum-Classical Interface} \\ Signal conversion and measurement readout};
\node[layer, below=of interface] (hardware) {\textbf{Quantum Chip / Physical Layer} \\ Qubits, native gates, decoherence sources};

\end{tikzpicture}
\caption{The QC stack, illustrating the layered flow from high-level applications to physical qubit execution. This diagram is based on the work of Fu et al. (2016)~\cite{10.1145/2903150.2906827} and Fu et al. (2017)~\cite{fu2017experimental}.}
\label{fig:qc-stack}
\end{figure*}

\subsection{Quantum Stack}\label{qc_stack}
The different components of a quantum computer can be described through a set of layers, resembling a hierarchy commonly used in classical computing~\cite{PhysRevX.2.031007, 10.1145/2494568, 8936946, 10.1145/2903150.2906827, fu2017experimental} (see Fig.~\ref{fig:qc-stack}). We will now briefly discuss the individual layers and components of such a quantum stack to provide a fundamental understanding.

\begin{enumerate}[label=(\alph*)]
\item \textbf{Application:}
The highest layer of the quantum stack comprises concrete applications where quantum algorithms are leveraged to solve real-world problems. The applications of quantum algorithms span across a variety of domains, commonly involving simulation, machine learning, and optimization tasks. In this layer, applications are specified at a conceptual layer, using abstractions of quantum algorithms as building blocks.

\item \textbf{Algorithm:}
The algorithmic building block required by the application layer is specified in a hardware-agnostic manner. In this layer, quantum algorithms are described formally, expressed through mathematical abstractions of quantum gates and subroutines. While the subroutines defined at this layer of abstraction can serve an application directly, such as the Grover algorithm~\cite{grover1996fast}, multiple subroutines are usually combined and wrapped by interfaces to tackle domain-specific problems. The Algorithm layer relies on deeper layers of the quantum stack to handle the physical implementation of primitive operations and to facilitate fault-tolerance. 

\item \textbf{Programming Paradigm and Languages:}
To express quantum algorithms, a system of notation is required. Some languages provide abstractions for data types representing quantum information science primitives and operations such as state preparation and qubit manipulation. Like classical languages, they can be grouped as either imperative, such as \texttt{Ket} ~\cite{da2021ket} and \texttt{Q\#} \cite{QsSpec2020} or functional, such as \texttt{LIQUi|>} ~\cite{wecker2014liquisoftwaredesignarchitecture} and \texttt{Quipper} \cite{Green_2013}. 
To facilitate QC experimentation, software development kits (SDKs), such as \texttt{Qiskit} ~\cite{qiskit2024}, \texttt{Cirq} \cite{Cirq} and \texttt{PennyLane} \cite{bergholm2022pennylaneautomaticdifferentiationhybrid} offer both classical simulation of quantum circuits and direct execution on real quantum processors, which can be accessed via cloud-based platforms. While error-correction usually occurs across lower layers, some SDKs provide error-mitigation features and often have a target domain, such as \texttt{PennyLane} for quantum machine learning~\cite{bergholm2022pennylaneautomaticdifferentiationhybrid}.

\item \textbf{Compiler, Runtime, and Arithmetic:}
For every programming language, a compiler is required to convert the abstract code specifying the algorithm into a lower-level set of instructions executable on a quantum processing unit (QPU). Furthermore, compilers ensure compatibility between the generated set of instructions and the hardware used for the execution. For instance, complex gates, such as the Toffoli gate, are commonly used in quantum algorithms, but are rarely present in the native gate set. Such gates need to be decomposed into natively available gates, for instance, transforming a Toffoli gate into a composition of $H$, $T$, $T^\dagger$ and $CNOT$ gates. 
Additionally, one must consider the limitations on qubit connectivity when performing operations between distant qubits. Different qubit realizations often require specific topological arrangements for qubits, such as the heavy-hexagonal lattice used by IBM's Falcon and Heron superconducting processors \cite{abughanem2024ibm}, where qubits are connected to at most three direct neighbors. On top of this, as in classical hardware, the runtime environment is responsible for tasks that must be performed during execution of the instruction set, such as the correct allocation of qubits~\cite{chong2017programming}. Finally, classical arithmetic units are adapted to the constraints of quantum information science, such as the requirement of reversible operations~\cite{thomsen2010reversible}.

\item \textbf{Instruction Set Architecture:}
Quantum instruction sets provide an abstraction layer between quantum software and hardware, specifying algorithms in terms of operations that can be executed on the QPU while hiding hardware implementation details. These sets are close to classical intermediate representation and assembly languages, describing all available operations on the device, including gates supported by the hardware, measurement operations, control flow instructions, and timing primitives.
Different instruction set architectures are often optimized for specific hardware types, such as the \texttt{Blackbird} quantum assembly language~\cite{Killoran_2019} for photonic processors, while others are completely hardware-agnostic, such as \texttt{OpenQASM}~\cite{cross2017openquantumassemblylanguage} and \texttt{cQASM}~\cite{khammassi2018cqasmv10commonquantum}.

\item \textbf{Microarchitecture:}
The micro-architecture layer translates the operations defined by the quantum instruction set into an execution pipeline of concrete physical control sequences. This layer orchestrates instruction decoding, timing, sequencing, and execution on the QPU processes~\cite{fu2017experimental}. Additionally, quantum algorithms are often hybrid in nature, containing both classical and purely quantum subroutines. The micro-architecture layer is also responsible for the scheduling and coordination between classical and quantum subroutines.

\item \textbf{Quantum to Classical:}
This interface layer manages the communication between classical abstraction layers and quantum hardware~\cite{reilly2015engineering}. It converts digital instructions from higher layers into analog control signals required by the quantum processor to implement quantum operations and measurements. Conversely, it processes analog signals from the quantum hardware, such as measurement outcomes, into digital data for higher-level layers. This layer is technology-dependent and ensures that the physical hardware's behavior aligns with logical operations above it. The quantum-to-classical conversion is challenging, as electronic interfaces capable of converting digital signals into analog pulses must be specially engineered to operate at cryogenic temperatures commonly required for quantum hardware. 

\item \textbf{Quantum Chip:}
The lowest layer of the stack is the quantum chip, or physical layer, which includes the hardware components of a quantum computer: physical qubits, single- and multi-qubit gates, measurement devices, and the host system. The host system, as specified in the alternative layer model from Ref.~\cite{PhysRevX.2.031007}, refers to a specifically engineered environment that supports computing with physical qubits of choice. Finally, in addition to these components, on a conceptual level, direct sources of noise and decoherence are also part of the physical layer of a quantum computer.
\end{enumerate}

\section{Related Work}\label{relatedwork}
Numerous studies have examined quantum benchmarking (QBM) techniques with the intention of structuring and comparing approaches in this diverse field~\cite{wang2022sok, acuaviva2024benchmarking, lorenz2025systematic}. In this section, we highlight studies that categorize, compare, and analyze various benchmarking techniques. 

A recently proposed high-level framework for quantum certification and benchmarking contextualizes these protocols based on three axes: complexity, information gain, and assumptions required~\cite{eisert2020quantum}. The technical review covers a broad range of common methods, making it particularly useful for gaining an initial overview of the field.

A complementary study by Resch et al.~\cite{resch2021benchmarking} examines QBM techniques through the lens of computer architecture and noise analysis. The authors categorize benchmarking approaches into three distinct classes: Qubit Benchmarking, (Quantum) Computer Benchmarking, and Fault Tolerant (Quantum) Computer Benchmarking. Qubit Benchmarking focuses specifically on the characterization of individual qubits and two-qubit systems, covering techniques such as quantum state tomography, quantum process tomography, and randomized benchmarking. In the second category, (Quantum) Computer Benchmarking, the authors focus on benchmarks which take a more holistic view on the system under observation and therefore attempt to evaluate the performance of the entire quantum system by executing specific quantum protocols. Finally, the chapter on Fault Tolerant (Quantum) Computer Benchmarking primarily offers an outlook and predictions of benchmarks for those devices rather than an extensive literature review. The authors attribute this to the limited research in the field, as fault-tolerant quantum devices have yet to be realized. They predict that in the post-NISQ era, benchmarks will likely focus more on real-world applications. 

The authors in Tomesh et al.~\cite{tomesh2022supermarq} propose a scalable, principle-driven quantum benchmark suite, \textit{SupermarQ}, designed to overcome key limitations of earlier benchmarks, such as the oversimplification of system performance into single-number metrics and the poor representativeness and scalability of synthetic benchmarks. To address these issues, they conducted an extensive literature review and categorized existing benchmarks, guiding the design of SupermarQ around diverse, meaningful, and scalable application workloads. The authors adopted the groups Gate-Level Characterization, Synthetic Benchmarks, Application Benchmarks, and Benchmark Suites. While this type of categorization sounds robust, it is deficient in technical explanation and lacks clear definitions. The rationale for each proposed category appears to be rooted in different principles, leading to potential ambiguity in practical application. While Gate-Level Characterization and Application Benchmarks are defined based on the intended use of the particular benchmark, Synthetic Benchmarks are defined by the technical paradigm they use. 

Another benchmark overview was proposed by Ref.~\cite{bharti2022noisy}, which divides them into four categories: randomized benchmarking, quantum volume, cross entropy benchmarking, and application benchmarks. This review focuses predominantly, though not exclusively, on benchmarks primarily designed for NISQ devices. 

A review of key techniques for near-term QC — including variational quantum algorithms, error mitigation, circuit compilation, and benchmarking protocols — introduces another categorization of quantum benchmarks~\cite{huang2023near}. The authors categorize the benchmarks at a high level based on whether they focus on gates or circuits. The gate-level benchmarking category differentiates between randomized benchmarking and cross entropy benchmarking, while circuit-level benchmarking includes random quantum circuit sampling, quantum volume and circuit layer operations per second, mirror circuits, and application-oriented benchmarks. While the paper offers a well-arranged high-level structure, it does not provide clear definitions. A more in-depth discussion of the various categories in terms of their application, usefulness, and dependencies, as well as an in-depth analysis of the shortcomings of benchmarking techniques, remains partially open.

Proctor et al.~\cite{proctor2025benchmarking} discuss the role of benchmarking in QC, highlighting the shift from low-level benchmarks like randomized benchmarking to high-level ones such as quantum volume, which assess entire systems. They emphasize the need for better-designed benchmarks, as current ones focus on isolated performance aspects and fail to capture the complexity required for achieving quantum utility. Six attributes are proposed for effective benchmarks: being well-motivated, well-defined, implementation-robust, system-robust, efficient, and technology-independent. Existing benchmarks are categorized into (1) computational problem, (2) compiler, (3) high-level program, and (4) low-level program benchmarks. The paper concludes with an outlook on improving benchmarks to better evaluate quantum systems and their potential to solve meaningful, real-world problems.

Hashim et al.~\cite{hashim2024practicalintroductionbenchmarkingcharacterization} provide a comprehensive and detailed overview of quantum characterization, verification, and validation (QCVV). In their work, they summarize the key underlying models and concepts essential for understanding QCVV. Additionally, they present an overview of various benchmarking techniques, including tomographic reconstruction methods, randomized benchmarks, holistic benchmarks, and fidelity estimation techniques. While their review is comparatively long and in-depth, offering a thorough introduction to the subject, it differs from our review in that it is not a systematic literature review.

In addition to these proposed frameworks and categorizations of QC benchmarks, there is also literature on specific subcategories, such as randomized benchmarking and its related frameworks~\cite{helsen2022general}.

Overall, existing benchmarking reviews have made valuable contributions by organizing subsets of the field, introducing initial taxonomies, and identifying challenges. However, these works vary considerably in scope, level of technical detail, and consistency in definitions, making it difficult to form a unified picture of QC benchmarking as a whole. Most prior reviews either focus on a single class of benchmarks or address only a few selected subcategories, which limits their ability to compare approaches across the entire QC stack or to highlight their interdependencies. Furthermore, the lack of standardized terminology and systematic methodology has led to overlapping categories and ambiguities in practical application.

Our work addresses these shortcomings by conducting the most comprehensive systematic literature review to date on QC benchmarking, covering the full range of hardware-, software-, and application-level approaches. We introduce a unified taxonomy aligned with the quantum stack, provide precise definitions for each category, and analyze their interconnections, enabling a more coherent and standardized view of the field.

\section{Methodology}\label{methodology}
\begin{figure*}
    \centering
    \includegraphics[width=\textwidth]{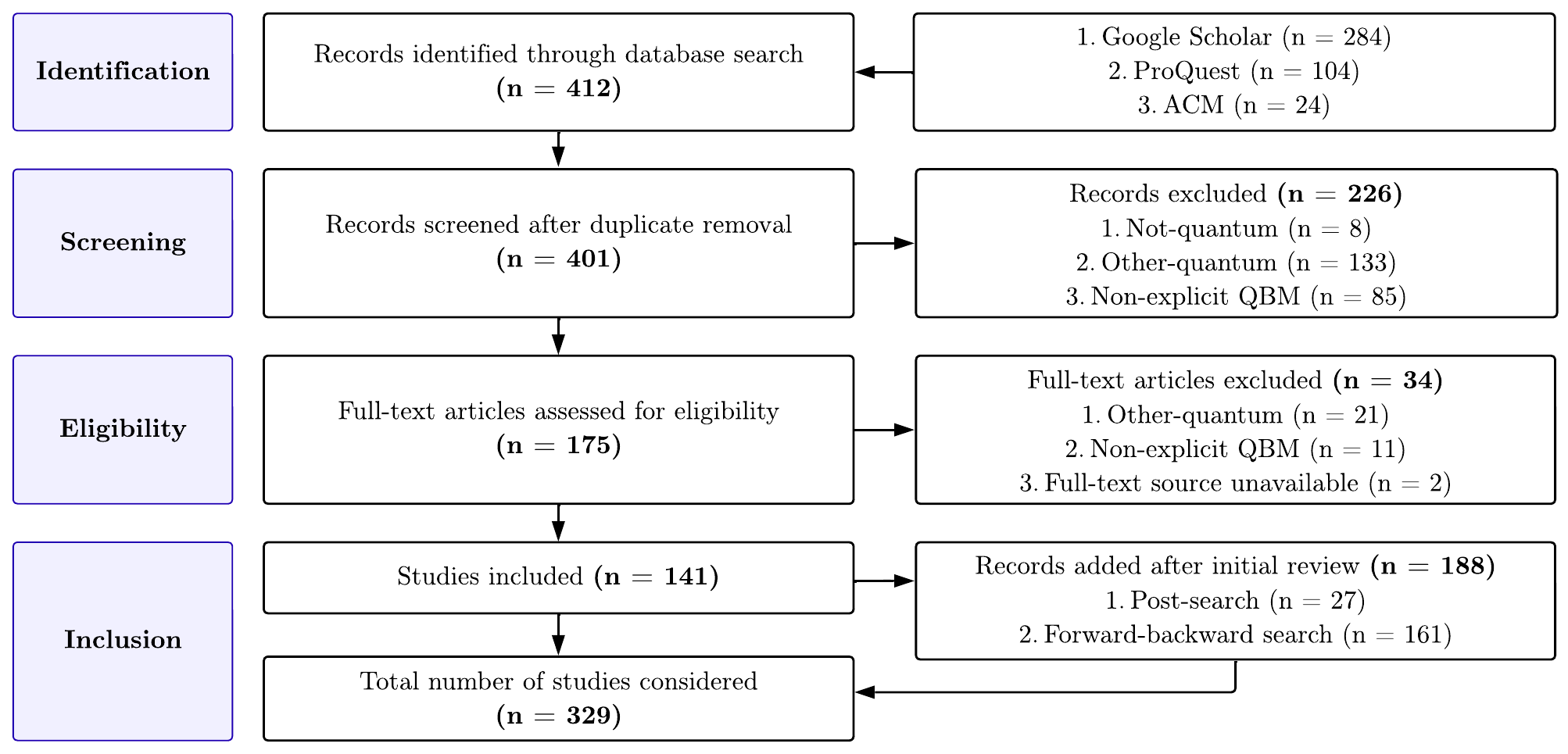}
    \caption{This flowchart illustrates the systematic literature review process, detailing the number of records that were identified, screened, and selected through both database searches and forward-backward search strategies.}
    \label{fig:literature-flow}
\end{figure*}

To remedy the discussed shortcomings in existing literature and provide profound definitions, this study is designed as a systematic literature review, adhering to the general guidelines of Xiao et al.~\cite{xiao2019guidance}, which emphasizes a structured and transparent review process, systematic quality assessment of sources, and ensured reproducibility through independent evaluations. To operationalize these principles, we refined our research protocol based on the systematic review template proposed by Biolchini et al.~\cite{biolchini2005systematic}. Our study can be classified as a secondary study, as it consolidates and analyzes previously published benchmarks and benchmarking techniques~\cite{grant2009typology} within the field of QC. The subsequent sections will outline the research methodology, ensuring traceability and reproducibility to achieve our goal of conducting a rigorous and systematic literature review. The various steps can also be traced using Fig.~\ref{fig:literature-flow}. During the identification phase, we collected all articles meeting our search criteria in ACM Digital Library, Google Scholar, and ProQuest, resulting in $412$ initial studies. After removing $11$ duplicate items, we conducted the title-abstract screening, excluding $226$ items based on our three exclusion criteria. Subsequently, we reviewed the remaining $175$ full-text articles for final eligibility, of which $32$ met the exclusion criteria and $2$ were unavailable in full-text form. The process resulted in $141$ included studies, to which we added $27$ more recent studies found through a post-search and $161$ studies resulting from a forward-backward search, resulting in the final $329$ studies considered in this review. While all $329$ studies were insightful to outline our taxonomy, we only explicitly cited $273$ of the total considered studies. $56$ studies were not explicitly addressed and cited in this review because they made only a limited contribution to the central topics, overlapped significantly with earlier studies, or were not sufficiently novel. However, they were taken into account for the organization of our taxonomy, so we conclude that our review remains complete.

\subsection{Research Objective}
Our research aims to develop a comprehensive understanding of QC benchmarking techniques by systematically categorizing them into a structured taxonomy. For each category, we develop a definition, explain the methodology of associated benchmarks, introduce relevant literature, and provide contextual insights.

Beyond classification, we analyze the relationships between benchmark categories, identifying key differences, similarities, and interdependencies within the QC stack. This examination offers a clearer perspective on how benchmarking techniques complement or contrast with one another, ultimately guiding their appropriate selection and use.

\subsection{Search Strategy}

For initial literature identification, we used three databases:ACM Digital Library, Google Scholar, and ProQuest. The selection of ACM Digital Library and Google Scholar was informed by established guidance for systematic reviews in software engineering~\cite{brereton2007lessons}, while ProQuest was additionally included to ensure broader interdisciplinary coverage, capturing publications that might not be indexed in field-specific databases. An initial Boolean search string was developed by synthesizing keywords from papers identified as relevant to our study focus. The following search formulation was then tested and refined in a trial run:
\\
\\
\textit{ABSTRACT = (``quantum'' AND (``comput*'' OR ``hardware'' OR ``system'' OR ``algorithm'' OR ``workflow'')) AND (``benchmark*'' OR ``comparison'' OR ``performance'' OR ``validation'' OR ``verification'' OR ``certification'')}
\\
\\
This trial revealed that certain keyword combinations frequently appeared in unrelated literature, necessitating adjustments to improve search relevance. Specifically, the combination \textit{``algorithm'' AND (``comparison'' OR ``performance'' OR ``validation'')} was common among articles in diverse areas with no relation to QC benchmarking. Additionally, the keywords \textit{``validation''} and \textit{``verification''} yielded some relevant results for low-level benchmarks but primarily retrieved studies on formal verification methods for programs on quantum hardware. Finally, we removed \textit{``certification''} from the search string, as relevant studies were retrievable without explicitly including it, and its inclusion did not improve search relevance. 

After these adjustments, we arrived at the final search string:
\\
\\
\textit{ABSTRACT = ``quantum'' AND (``comput*'' OR ``hardware'' OR ``system'' OR ``workflow'') AND (``benchmark*'') }
\\
\\
The final search string targeted papers where the abstract included the term \textit{``quantum''} to ensure relevance to quantum technologies, combined with general system-level terms \textit{(``comput*'', ``hardware'', ``system'', ``workflow'')} to capture both software and hardware benchmarking contexts. The asterisk symbol (*) in the search string denotes a wildcard, allowing for the retrieval of all word variants, such as \textit{``compute''}, \textit{``computing''}, \textit{``computer''}, or \textit{``benchmark''} and \textit{``benchmarking''}. The addition of \textit{``benchmark*''} ensured a focus on performance evaluation and comparative studies, while excluding broader fields such as algorithm validation or formal verification.

During our search, we encountered some paywalled publications. In these cases, we documented the references and assessed their relevance at a later stage. In our later analysis, we found no significant negative effects on our results or any directional bias. After conducting the systematic literature search, we supplemented the literature we included by performing forward-backward searches to address any potential gaps in the systematic search.

\subsection{Inclusion and Exclusion Criteria}
Our study primarily includes peer-reviewed articles and papers published between January 2000 and June 2025. Literature from other reputable sources that were not necessarily subject to a peer review process, such as technical reports, white papers, or dissertations, was taken into account provided that it was generally accepted and relevant to our study.
To be included, literature must fulfill all of the following criteria: 
\begin{itemize}[nosep]
    \item propose a new benchmark or provide a comprehensive description of an existing benchmark, 
    \item assess the quality or capabilities of one or more parts of the QC stack or workflow, 
    \item either apply the benchmark or provide a theoretical analysis of its performance. 
\end{itemize}
\medskip

We also applied exclusion criteria to ensure a focused selection of relevant literature and to avoid unnecessary complexity. If a paper met any of the following exclusion criteria, it was excluded from the review, even if it otherwise satisfied the inclusion criteria: 
\begin{itemize}[nosep]
    \item literature not written in English,
    \item literature with a non-explicit focus on QC benchmarking (e.g. merely mentioning or indirectly addressing),
    \item benchmarks about other quantum technology related types (e.g. quantum sensing or quantum communication),
    \item non-universal QC.
\end{itemize}

We excluded non-universal QC platforms, such as analogue quantum simulators or quantum annealers, as they operate under different computational models and lack general-purpose programmability, making them fundamentally distinct from the gate-based architectures targeted in this review. 

Initially, the collected literature was screened in a double-blind setting. During this step, two of the authors of this review independently considered the title and abstract of every collected study, applying the inclusion and exclusion criteria. Search results that did not meet all the inclusion criteria or met at least one exclusion criterion were categorized accordingly. If both reviewers independently agreed on the labeling, the article was either considered in the next screening step or excluded. In case of disagreement, a discussion was held until unanimity was reached. In case of duplicate studies, we only include the most recent and/or comprehensive study. 

\subsection{Study Selection \& Data Synthesis}
Following the initial title and abstract screening, we conducted a thorough full-text review of the collected studies. During this in-depth review, we evaluated the quality, relevance, and robustness of each study, reassessing if the studies met any exclusion criterion. This involved assessing the rigor of the study design, the validity of the findings, and the clarity and transparency of the reporting. 

To create an initial overview of the collected literature, a statistical approach was performed using the technique of topic modeling, creating clusters of related studies. In particular, the BERTopic~\cite{grootendorst2022bertopic} toolset was used, performing a set of transformations including document embedding, dimensionality reduction, clustering of reduced embeddings, and automatic cluster labeling. Different techniques and models were used for each transformation, discussed in detail in Appendix~\ref{BERTopic}. The resulting clusters served as a first categorization of the collected literature, which we later refined through an in-depth inspection of the clusters and several feedback rounds. Finally, we complemented the refined clusters through a forward-backward search. While all collected studies were included in our clustering, only important studies that are truly relevant to the subject area are considered explicitly in this review. This is intended to improve the compactness of the review without undermining its relevance and completeness.

The entire research methodology, as well as its execution, was monitored throughout the process by a researcher unfamiliar with QC but with experience in literature analysis. 

\section{Results}\label{results}
\begin{figure*}[htbp]
    \centering
    \includegraphics[width=\textwidth]{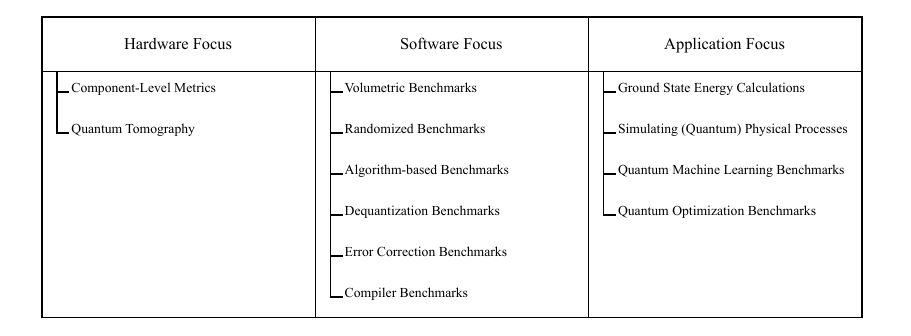}
    \caption{Hierarchical classification of QBM techniques. The figure illustrates the primary categorization of benchmarks based on their alignment with the QC stack, while within each top-level category, benchmarking techniques are further grouped by their methodological principle. Each category corresponds to one or several different layers of the QC stack and stakeholder interests -- ranging from low-level device characterization to high-level application performance. This enables structured comparisons of their technical principles, objectives, and scopes. In our review we further subdivided the subcategories where appropriate.}
    \label{fig:benchmarking_hierarchy}
\end{figure*}

To systematically structure the field of QC benchmarking, we introduce a hierarchical categorization that reflects both the primary users of benchmarks and the specific aspects of system performance they evaluate. This structured approach enhances clarity in the field and aligns benchmarking techniques with the needs and interests of different stakeholders within the QC ecosystem.
At the first level, benchmarking techniques are categorized based on their primary user groups and the specific aspects of system performance they evaluate, following a natural hierarchy akin to the QC stack. This hierarchical organization is designed to guide readers through the extensive and detailed landscape of benchmarking techniques, allowing them to efficiently identify relevant methods based on their interests or roles within the QC ecosystem. Reflecting the layered structure of the QC stack, this classification consists of three principal categories:

\begin{enumerate}
    \item \textbf{Hardware Focus} – corresponding to the lower levels of the QC stack, targeting people interested in single hardware characteristics and components of the quantum system, rather than the interplay of components and the joint functionality of the system. These are often hardware developers, physicists, or engineers, people interested in the particular component at hand.
    \item \textbf{Software Focus} – aligned with intermediate layers of the QC stack, evaluates system-wide performance, focusing on the interaction of hardware components and software execution, typically independent of specific end-user applications. This focus primarily serves algorithm and software developers designing and testing quantum algorithms and frameworks.
    \item \textbf{Application Focus} – positioned at the top of the QC stack, addressing the needs of end users who execute quantum applications for specific tasks, such as optimization or simulation. These benchmarks assess how well a quantum system performs a specific computational application, prioritizing task-specific performance over general hardware or software capabilities.
\end{enumerate}

This categorization mirrors the QC stack, where hardware benchmarks operate at the lower layers, software-related benchmarks at the intermediate layers, and user-driven benchmarks at the application layer of the QC stack. 
In the next hierarchical level beneath these three categories, benchmarks are further grouped by their underlying methodology, clustering techniques that share similar logical or technological principles. This secondary classification improves comparability, making it easier to assess similarities and differences in benchmarking approaches.

\subsection{Hardware Focus}\label{hardware}
The Hardware Focus category encompasses benchmarking techniques primarily designed for stakeholders engaged in the research and development of quantum hardware components and their underlying physical properties. The target group for this category includes physicists and engineers working at the lowest levels of the quantum stack, focusing on understanding and improving individual hardware elements.
Benchmarks in this category provide hardware-centric metrics that evaluate separately the fundamental building blocks of quantum computers, such as qubits, quantum gates, and control pulses. Their physical properties --including coherence times, gate fidelities, error rates, and crosstalk effects-- directly influence quantum computation but do not fully determine the overall computational performance of the system. By isolating and analyzing these properties in detail, hardware benchmarks enable the precise characterization and optimization of individual components. 
This hardware-focused benchmarking section spans from component-level and operation speed metrics to tomographic techniques, providing complementary perspectives for quantifying quality, performance, and limitations at the lowest layers of the quantum stack.

\subsubsection{Component-Level Metrics}

\begin{figure}[htbp]
\centering
\begin{tcolorbox}[title=BOX I, coltitle=white, colbacktitle=blue!75!black, sharp corners=south]
\textbf{Overview of Component-Level Metrics} 
\vspace{3pt} 

Standard metrics are useful to concisely describe different properties of a QPU, mirroring its quality. In contrast to higher-level metrics, such as \textbf{Quantum Volume} \cite{bishop2017quantum}, \textbf{CLOPS} \cite{wack2021qualityspeedscalekey} or \textbf{Q-score\textsuperscript{TM}} \cite{martiel2021benchmarking}, component-level metrics directly describe the physical properties and limitations of qubits and quantum gate implementations, as well as the effect of noise on the system, instead of quantifying the aggregated quality of the entire system. 

The following component-level metrics will be described in the subsequent section:

\begin{enumerate}
    \item \textbf{Primary Hardware Properties}
    \begin{enumerate}
        \item Number of qubits
        \item Connectivity topology 
        \item Native gate set
    \end{enumerate}
    \item \textbf{Qubit Quality Metrics}
    \begin{enumerate}
        \item $T_1$ energy relaxation time
        \item $T_2$ qubit dephasing time
        \item \textit{Q}-factor
    \end{enumerate}
    \item \textbf{Operation Quality Metrics}
    \begin{enumerate}
        \item Process fidelity
        \item Average gate (in-)fidelity
        \item $r$ (Output of the randomized benchmarking protocol)
        \item Diamond norm
        \item Readout fidelity
    \end{enumerate}
    \item \textbf{Operation Speed Metrics}
    \begin{enumerate}
        \item Gate execution time
        \item Measurement time
        \item Qubit reset time
    \end{enumerate}
\end{enumerate}
\vspace{5pt} 

Further reviews of these and multiple additional metrics to characterize different levels of the QC stack can be found in \cite{lall2025review, van2023evaluating, gilchrist2005distance}.

\end{tcolorbox}
\end{figure}

\paragraph{Primary Hardware Properties} 

\noindent\makebox[\columnwidth]{%
    \fbox{%
        \parbox{0.95\columnwidth}{%
        The number of qubits, the way these qubits are interconnected, and the set of natively supported operations, are among the main properties characterizing a QPU.
        }%
    }%
}
\\

The \textbf{number of usable qubits} is the first primary hardware characteristic determining which algorithms can be mapped on a device, as well as posing a constraint on the maximum dimensionality of outputs that can be encoded in its qubits. An equally relevant property for the characterization of the device is the structure of connectivity between qubits, known as the hardware \textbf{topology}, which can greatly vary depending on the physical implementation of the QPU~\cite{linke2017experimental}. When implementing an algorithm, there is often a mismatch between the algorithm specification and the available topology. For example, if multi-qubit gates between non-neighboring qubits are specified, additional SWAP operations are required, increasing complexity and error rates. The third primary hardware property of a QPU is given by its \textbf{native gate set}, the set of operations that can be physically executed on the device. Not every native gate set allows universal quantum computation~\cite{shi2002both}, which requires an appropriate combination of gates to approximate any unitary operation on the system's qubits, including operations beyond the Clifford group. Furthermore, most quantum algorithms are specified with gates outside the native gate set. As a result, the algorithmic complexity depends on the effort required to compile natively available gates into the gates required by the algorithm.

\paragraph{Qubit Quality Metrics} 

\noindent\makebox[\columnwidth]{%
    \fbox{%
        \parbox{0.95\columnwidth}{
        The quality of qubits is determined by their ability to maintain a desired state for a sufficient duration, characterized by coherence times.
        }%
    }%
}
\\
 
Coherence in quantum systems can be characterized by three primary timescales: the relaxation time $\mathbf{T_1}$ and the dephasing times $\mathbf{T_2}$ and $\mathbf{T_2^*}$. The $T_1$ time describes energy relaxation, quantifying the timescale over which a qubit decays from the excited state $\ket{1}$ to the ground state $\ket{0}$. The probability of finding the qubit in the excited state $\ket{1}$ decays exponentially with time and can be fitted to an exponential decay:
\begin{equation}
    P_{|1\rangle}(t) \propto e^{-t/T_1},
\end{equation}
where $t$ denotes the time variable. 

Complementarily, the $T_2^*$ time quantifies the timescale over which a qubit maintains phase coherence in superposition states. For example, a qubit initialized in the state $\ket{+} = \frac{\ket{0} + \ket{1}}{\sqrt{2}}$ may, due to dephasing, evolve into a statistical mixture of $\ket{+}$ and $\ket{-}$, making it impossible to determine its original phase relationship. One method to estimate $T_2^*$ is the Ramsey experiment, after the work from Ref.~\cite{PhysRev.78.695}. Here, a qubit is prepared in a superposition state $\ket{+}$ through a first $\pi/2$-Pulse, and, following a waiting period $t$, a second $\pi/2$-Pulse is applied, after which a final measurement in the computational basis is conducted. The decay time $T_2^*$ is obtained by fitting an exponential decay to observations obtained from different waiting times $t$:
\begin{equation}
    P_{\ket{1}}(t) \propto e^{-t/T_{2}^*}.
\end{equation}

Finally, the Hahn echo~\cite{PhysRev.80.580} $T_{2}$ – the spin-echo coherence time – is the time over which a qubit’s phase coherence decays when a refocusing $\pi$ pulse (Hahn echo sequence) is used to cancel quasi-static dephasing noise. In a Hahn echo experiment, one observes an exponential decay of the qubit’s coherence with time constant $T_{2}$ (fitting the measured signal to $\propto e^{-t/T_{2}}$). Because the echo removes inhomogeneous broadening from low-frequency noise, the Hahn-echo is longer than the free-induction dephasing time $T_{2}^{*}$ obtained from a Ramsey experiment. However, $T_{2}$ remains fundamentally limited by energy relaxation: any $T_{1}$ decay process also causes loss of phase coherence. In fact, one finds 
\begin{equation}
    \frac{1}{T_{2}} = \frac{1}{2T_{1}} + \frac{1}{T_{\varphi}}
\end{equation}
with $T_{\varphi}$ the pure dephasing time, which implies the inequality $T_{2} \le 2T_{1}$, with equality only in the absence of any extra dephasing noise.

Finally, the \textit{Q}-\textbf{factor}, closely related to dephasing times, measures how many operations can be performed in a single $T_2^*$ time, averaged over all qubits~\cite{van2023evaluating}. This gives a more intuitive reference of how many gates can be executed while the qubits retain the necessary phase information.

\paragraph{Operation Quality Metrics} 

\noindent\makebox[\columnwidth]{%
    \fbox{%
        \parbox{0.95\columnwidth}{%
        The quality of quantum gates is generally quantified in terms of a similarity metric between the implementation of a quantum gate and the idealized mathematical operation it should replicate.
        }%
    }%
}
\\

An intuitive measure of the fidelity of a gate is the trace distance between its process matrix and the unitary corresponding to its ideal operation. The following \textbf{process fidelity} formulation, as presented in Ref.~\cite{korotkov2013error}, as well as other distance measures to compare real and ideal quantum processes, can be derived from Ref.~\cite{gilchrist2005distance}:  
\begin{equation}
\label{eq:process_fidelity}
F_{\text{process}}(U,\mathcal{E}) = \operatorname{Tr}\left(\chi^U \chi^{\mathcal{E}}\right),
\end{equation}
where \( \chi^U \) is the Choi process matrix of the ideal unitary operation \( U \) and \( \chi^{\mathcal{E}} \) is the Choi process matrix of the implemented quantum operation \( \mathcal{E} \). It is assumed that the implemented operation is trace-preserving. In a more general setting, where the ideal operation is not unitary, this definition should be replaced by the \textbf{Uhlmann fidelity}:
\begin{equation}
    \label{eq:uhlmann_fidelity}
    F_\text{Uhlmann}(U,\mathcal{E}) = \left( \operatorname{Tr} \sqrt{ \sqrt{\chi^U} \, \chi^\mathcal{E} \, \sqrt{\chi^U}} \right)^2.
\end{equation}

This expression is state independent but expensive to calculate, as the process matrix of the characterized gate is obtained through process tomography. However, recent methods greatly reduce the cost of fidelity estimation, such as Direct Fidelity Estimation~\cite{flammia2011direct, zhang2021direct} and Shadow Tomography~\cite{Aaronson_shadow} by circumventing the full reconstruction of the process matrix. In contrast to the process fidelity metric, the \textbf{average gate fidelity}~\cite{schumacher1996sending, nielsen2002simple} measures the average state fidelity between the output of the characterized gate and the output of the ideal operation for all possible inputs (in practice for a large number thereof):
\begin{equation}
    \label{avg_fidelity}
    F_\text{average}(U, \mathcal{E}) = \int d\psi \bra{\psi} U^\dagger \mathcal{E}(\ket{\psi}\bra{\psi}) U |\psi \rangle.
\end{equation}

It is possible to derive the average gate fidelity from the process fidelity and vice versa~\cite{horodecki1999general, nielsen2002simple}:
\begin{equation}
    \label{avg_process_relation}
    F_\text{average}(U, \mathcal{E)} = \frac{dF_\text{process}(U,\mathcal{E}) + 1}{d + 1},
\end{equation}

where $d$ denotes the dimension of the channels. The \textbf{average gate infidelity} (AGI), given by $1 - F_\text{average}$, was believed to correspond to the output of the randomized benchmarking protocol $\textbf{r}$. However, AGI was later demonstrated to depend on the representations used for the imperfect and ideal gates~\cite{proctor2017randomized}, while the error metric $r$ is independent of this representation choice, making $r$ a more generalizable metric.  

Usually, the metric $r$ is referred to as \textbf{gate fidelity} in industry reports, as it underestimates AGI and accounts for average gate performance over all input states rather than worst-case performances~\cite{proctor2017randomized}, in contrast to the \textbf{diamond norm} metric~\cite{aharonov1998quantum}, which quantifies the worst-case error of a gate and can be calculated numerically~\cite{benenti2010computing}. Using $r$, both single-qubit gate errors and two-qubit gate errors are usually reported in industrial QPU specifications~\cite{blogMeetWillow}, as well as the \textbf{readout fidelity} or the \textbf{measurement error} metric.  

\paragraph{Operation Speed Metrics} 

\noindent\makebox[\columnwidth]{%
    \fbox{%
        \parbox{0.95\columnwidth}{%
        Speed metrics quantify the time required to execute the central operations of a QPU, including the initialization and measurement of qubits, as well as the application of quantum gates. 
        }%
    }%
}
\\

While a large effort in the development of QPUs goes towards improving the quality of qubits and the fidelity of gates, reducing the time required to execute operations is equally important in order to obtain useful hardware, as sufficiently fast operations are required to compute results within the coherence time of qubits. Moreover, a quantum algorithm which requires less computational steps than a classical alternative will only be faster on a QPU if the time required to run every operation is less than the time required to simulate the circuit classically. Finally, evidently, the duration of individual operations specific to a QPU determines the time required to execute more complex algorithms.

The first speed metric is the \textbf{gate execution time}. The time to execute different gates is highly dependent on the hardware platform, and in some cases, some gates do not need an actual pulse and can be executed by modifying the parameters of the following gates~\cite{mckayPhysRevA.96.022330}. For this reason, to realistically quantify gate execution times, the \textbf{time taken to perform a general single-qubit gate} on the device is the metric of choice. This provides an estimate of the average execution time over all possible single-qubit gates, as a general unitary can be decomposed into sequences of $R_z$ and $R_x$ gates. Other relevant operations which contribute to the total time required to execute a quantum circuit are the \textbf{readout} or \textbf{measurement time}, as well as the time required to reset the state of qubits after the execution of a circuit, known as the \textbf{reset time}. All of these metrics are highly dependent on the hardware platform and the natively available operations~\cite{arute2019quantum, nunnerich2024fast}.

\subsubsection{Quantum Tomography} 
\tcbset{
  colback=blue!5!white, colframe=blue!75!black, fonttitle=\bfseries,
  coltitle=black, boxrule=0.75pt, width=\columnwidth, arc=2mm
}

\begin{figure}[htbp]
\centering
\begin{tcolorbox}[title=BOX II, coltitle=white, colbacktitle=blue!75!black, sharp corners=south]
\textbf{Overview of Tomographic Techniques}

The term ``Tomography'' comes from the Greek ``tomos'' -- ``part'' and ``graphein'' -- ``to write''. In the context of QC, Quantum Tomography refers to approximatively acquiring the state of a quantum physical system or the operator governing its time evolution, based on cross-sections of the system. This process is conducted by repeatedly measuring identically prepared systems, forming frequency counts, which are then used to infer probabilities used to create a mathematical model of the characterized system. It enables the detailed characterization of quantum systems by inferring their density matrix or process matrix from experimental observations. 

While tomography provides deep insights into quantum hardware performance, its scalability is a major challenge, as the number of measurements required to accurately model quantum states and processes grows exponentially with the system size. This makes efficient and approximative tomographic methods crucial for practical applications in larger quantum systems. An overview about measurement and sample complexities of selected QST and QPT variants is provided in Tab.~\ref{tab:quantum_tomography_methods}. Quantum Tomography is typically classified according to the subject of characterization. In our review, we will take a closer look at the following techniques:
\vspace{0.5em} 
\begin{itemize}[nosep]
    \item Standard State and Process Tomography
    \item Gate Set Tomography
    \item Structure-Exploiting Tomography
    \item Targeted Information Tomography
    \item Statistical and Machine Learning Tomography
\end{itemize}

\end{tcolorbox}
\end{figure}

\definecolor{deftitle}{RGB}{27, 53, 133}
\definecolor{defgreenborder}{RGB}{60, 179, 113}

\newcounter{defcounter}
\setcounter{defcounter}{1}

\newtcolorbox{defbox}[2][]{%
    colback=white,
    colframe=black,
    sharp corners, fonttitle=\bfseries\normalsize\rmfamily,
    colbacktitle=white,
    coltitle=deftitle,
    center={yshift=-0.25mm},
    before upper={\stepcounter{defcounter}}, 
    title={\textit{1.\thedefcounter:} #2},
}

\paragraph{Standard State and Process Tomography} 

\noindent\makebox[\columnwidth]{%
    \fbox{%
        \parbox{0.95\columnwidth}{%
            Quantum Tomography seeks to reconstruct mathematical representations of quantum states and processes from a set of measurements of the same underlying system.
        }%
    }%
}
\\

\textbf{Quantum State Tomography} (QST) encompasses methods for estimating reconstructions of quantum states in the form of a quantum wave function~\cite{Torlai2018}, a state vector~\cite{lange2023adaptive}, a Wigner function~\cite{leonhardt1996discrete, he2024efficient}, a density matrix (DM)~\cite{Schmied_2016, binosi2024tailor} or any other representation~\cite{guo2024quantum} of a quantum state. On the other hand, \textbf{Quantum Process Tomography} (QPT) encompasses methods for estimating the reconstruction of any quantum process that describes the evolution of a quantum system, such as a quantum gate or a noise channel. Process estimations take the form of a completely positive and trace-preserving map through a convenient representation, such as a Choi matrix or a Kraus representation~\cite{10.5555/2871422.2871425}. In addition to QST and QPT, \textbf{Detector- or Measurement Tomography} is sometimes considered as a separate branch in the literature~\cite{blumoff2016implementing}, but it can also be seen as a generalization of QST~\cite{nielsen2021gate_set}.

The first step of quantum tomography consists in gathering information about the system to be studied. For QST, data is collected by performing a set of measurements on identically prepared states of the quantum system. Due to the destructive and stochastic nature of quantum measurement, multiple copies of the same quantum state must be created and measured to model it with the desired accuracy. In the case of QPT, data is collected by preparing identical copies of an initial state, which are subsequently evolved by the process to be characterized. The resulting states are measured again in a desired basis in order to estimate the effect of the investigated process. These measurements provide the necessary data to reconstruct models of the underlying states and processes using a variety of methods. The first theoretical proposals of QST~\cite{vogel_theoretical_concept} and QPT~\cite{chuang1997prescription} focus on the complete reconstruction of states and processes. Since the first experimental attempts to obtain such reconstructions~\cite{ashburn1990experimentally, smithey1993measurement, leonhardt1996discrete}, multiple variants of QST and QPT have been developed. Although earlier methods were restricted to small systems, ranging from one to three qubits~\cite{resch2005full}, more recent methods are able to characterize much larger systems, such as a 12-qubit state on a quantum device~\cite{hu2024experimental} and a 20-qubit state in a simulation~\cite{kurmapu2023reconstructing}. Modern tomography methods greatly decrease the required number of measurements and/or the computational cost of calculating the estimate from these measurements. For example, $531 \, 441$ ($= 3^{12}$) observable expectation values are required for a conventional full QST of a \textit{W} state over 12 qubits, the QST method from Ref.~\cite{hu2024experimental} achieves similar results with only 243 parallel expectation value estimations, greatly decreasing measurement complexity. Other tomography variants, such as Ref.~\cite{nielsen2021gate_set} focus on improving the quality of tomographic reconstructions rather than the computational and experimental efficiency of the tomographic process.

Furthermore, various alternative methods for QST were transferred to the domain of QPT through the Choi–Jamiołkowski isomorphism~\cite{choi1975completely, jamiolkowski1972linear, de1967linear}, describing the correspondence between quantum states and channels~\cite{jiang2013channel}.

\paragraph{Gate Set Tomography} 

\noindent\makebox[\columnwidth]{%
    \fbox{%
        \parbox{0.95\columnwidth}{%
            Gate Set Tomography (GST) extends QST and QPT by simultaneously characterizing all native gates, including state preparation and measurement (\textit{SPAM}) operations. In contrast to QPT, GST does not assume any component (including \textit{SPAM} operations) to be perfectly calibrated. Thus, by simultaneously characterizing the entire gate set as a collection of process matrices, GST can directly identify and quantify \textit{SPAM} errors, commonly misattributed to errors of the characterized gates in QPT.
        }%
    }%
}
\\

Most QST and QPT methods estimate quantum states and processes based on assumptions about the fidelity of certain components of the underlying system. Particularly, the \textit{SPAM} operations used during the tomographic setup are assumed to be perfect. In practice, this is problematic, as beyond possible state decoherence and gate implementation errors that QST and QPT respectively account for, \textit{SPAM} operations often-times introduce errors. 
Thus, \textit{SPAM} errors can be misattributed to errors in the state or process being characterized, leading to systematic inaccuracies in traditional tomographic techniques. Building on ideas from~\cite{mogilevtsev2009relative, mogilevtsev2012self, merkel2013self_consistent}, the protocol of \textbf{GST}~\cite{blume2013robust} addresses these issues, removing any assumptions about the quality of \textit{SPAM} operations, simultaneously characterizing the entire set of gates. 

The input for GST is the underlying quantum gate set to be characterized, including \textit{SPAM} gates. The selected gates are arranged into probing sequences that are then executed to generate measurement outcomes. The GST outputs a self-consistent set of process matrices describing each component of the gate set, including all intermediate gates, state preparations, and measurements. GST is a powerful tool for evaluating the robustness of a system, as the process matrices obtained through this protocol reflect the influence of both coherent errors arising from miscalibrated control systems and incoherent errors arising from interactions with the environment. In contrast, traditional QPT might confound coherent errors arising from imperfect \textit{SPAM} operation implementations with incoherent errors. 

In spite of the high precision achieved by GST, it has certain limitations when probing larger systems. GST scales exponentially with the number of qubits, even polynomially faster than conventional QPT~\cite{greenbaum2015introduction}. This is due to the need to characterize not only the gates corresponding to the process to be characterized, but also the state preparation and measurement operations at the same time. Blume-Kohout et al.~\cite{blume2013robust} provide a formula for the required number of experiments $M$ for a gate set containing $K$ gates operating on $d$ ($d = 2^n$ for $n$ qubits) dimensions: $M \approx Kd^4 - (K-2)d^2 - 1$. 

To accurately estimate even a minimal complete gate set consisting of three single-qubit gates, around $80$ experiments are required. For a gate set consisting of four 2-qubit gates, more than $4000$ experiments are needed, compared to $16$ and $256$ experiments to estimate a single 1- or 2-qubit gate with QPT~\cite{greenbaum2015introduction}. In addition to possible scaling issues, the first implementations of GST lose precision when characterizing larger systems, as crosstalk between qubits is not generally captured. However, adaptations of GST have been introduced to address this problem~\cite{rudinger2021experimental}. Furthermore, modified versions of GST such as \textbf{Compressive GST}~\cite{brieger2023compressive} or Shadow estimations to obtain gate set properties~\cite{helsen2023shadow} considerably increase the efficiency of GST, making this protocol experimentally feasible.  Considering the ability of GST to characterize \textit{SPAM} errors and decouple their effect from intermediate gate imperfections, protocols of this type remain relevant to precisely characterize quantum devices.

\paragraph{Structure-Exploiting Tomography} 

\noindent\makebox[\columnwidth]{%
    \fbox{%
        \parbox{0.95\columnwidth}{%
            Characterized states and processes often have special structural properties such as high purity, symmetries or sparsity in a known basis. Structure-Exploiting Tomography incorporates known or anticipated structural information of this kind, reducing the required resources to create a mathematical reconstruction of the underlying state or process.
        }%
    }%
}
\\

One of the first approaches to reduce the computational cost of QST by making assumptions about the characterization target was \textbf{Compressed Sensing QST} (CS-QST). For states with a DM close to a matrix of rank $r$, where $r$ is small compared to the dimension of the system (low-rank states), CS-QST can achieve a quadratic reduction in the number of required measurement settings~\cite{gross2010compressedSensing}. The reconstruction of the DM of a low-rank state can be estimated using an efficient convex optimization algorithm~\cite{donoho2006compressed}. Furthermore, CS was employed to develop an experimentally efficient method for QPT, \textbf{Compressed Sensing QPT} (CS-QPT)~\cite{compressive_qpt}. In this case, if the process matrix is \textit{s-sparse}, meaning that it can be well approximated by a process matrix with at most $s$ non-zero elements in some known basis, only $\mathcal{O}(s\log d)$ measurement configurations are required, where $d$ is the dimension of the Hilbert space. The authors show that this condition is often fulfilled in practice~\cite{compressive_qpt}. CS-tomography has been extensively analyzed, with studies discussing error bounds and sample complexities~\cite{Flammia_2012_compressed_sensing_analysis} as well as experimental applications~\cite{riofrio2017experimental}. Recent advances in this space include \textbf{Objective Compressive QPT}, which fully characterizes non-full-rank, completely positive quantum processes~\cite{teo2020objective}, and the integration of compressive sensing into Gate Set Tomography~\cite{brieger2023compressive}.

A further structural property that has been exploited to develop more efficient tomographic schemes is permutational symmetry. \textbf{Permutationally Invariant (PI) tomography}~\cite{toth_pi_tomography} assumes that the characterized state remains unchanged under permutations of its subsystems. Thus, instead of reconstructing the full DM, PI tomography only estimates the PI part of the DM, requiring only a quadratically scaling number of measurements with increasing system size. Many widely used states, such as the \textit{GHZ} state, \textit{W} states and symmetric Dicke states, are PI, meaning that a reconstruction of the PI subspace of the density operator provides a very good estimation of the true state. DM reconstruction schemes tailored for PI tomography have been developed, taking advantage of the symmetry of PI states to efficiently store and process the states, resulting in the reconstruction of density matrices for states of up to 20 qubits~\cite{Moroder_pi_reconstruction}. Furthermore, CS-tomography can be conducted in the PI subspace, combining the strengths of both methods to further reduce the number of required measurement settings~\cite{schwemmer_experimental_comparison_pi_cs}.

Another direction for Structure-Exploiting Tomography are \textbf{Matrix Product States} (MPS) and \textbf{Finitely Correlated States} (FCS), two special representations, particularly well suited for pure states and one-dimensional systems. MPS and FCS representations can be obtained through methods requiring only a linear number of experimental operations and polynomial post-processing~\cite{Cramer2010}, making them feasible for larger systems, with experimental results achieving the characterization of 14-qubit states~\cite{lanyon2017efficient}. Furthermore, MPS representations are useful for systems that have locally correlated states~\cite{lanyon2017efficient}, e.g., 2D cluster states, which can be employed to implement any quantum algorithm. Alternative tensor network representations, such as Projected Entangled Pair States, also provide a useful way to compactly represent quantum states~\cite{Orus_tn_mps}. Recent advances in this direction integrate tensor networks with Shadow Tomography~\cite{Akhtar2023scalableflexible_cst} and the development of a QPT approach based on unsupervised learning and tensor networks~\cite{torlai2023quantum}. Finally, a MPS-based variational Ansatz was recently used to perform tomography of non-symmetrical states (more challenging to characterize than symmetric states such as \textit{GHZ} and \textit{W} states), achieving high-quality reconstructions with fast convergence on a 20-qubit simulator~\cite{kurmapu2023reconstructing}.

\paragraph{Targeted Information Tomography}

\noindent\makebox[\columnwidth]{%
    \fbox{%
        \parbox{0.95\columnwidth}{%
            Traditional QST and QPT methods create complete mathematical reconstructions of quantum states and processes. In contrast, Targeted Information approaches extract only practically relevant information about specific properties or performance metrics of the characterization target. 
        }%
    }%
}
\\

A clear example of Targeted Information Tomography is given by the method of \textbf{Direct Fidelity Estimation} (DFE)~\cite{flammia2011direct,da2011practical}. As the name suggests, instead of reconstructing a state or process representation, DFE aims solely at finding an estimate of the fidelity between a target state and the actual experimental implementation of this state. Although not explicitly defined as a form of tomography, DFE follows the same structure of QST and QPT approaches, capturing information about the characterization (or in this case, more precisely, certification) target based on measurements of multiple copies thereof. Carefully selecting the Pauli measurement operators that are most likely to detect deviations from the desired state, it is possible to obtain a fidelity estimate with a constant number of distinct Pauli measurement observables up to a constant additive error~\cite{flammia2011direct}. However, each of the distinct measurements must be repeated multiple times to estimate the corresponding expectation value. In the best case (true for stabilizer states), a \textit{constant} number of repetitions is enough but in the worst case, $2^n$ measurements are needed for an $n$-qubit system. A recent approach, which treats fidelity estimation as a machine learning classification task, achieves high quality estimations with a constant number of measurements for any kind of state, at the cost of training a neural network for this task~\cite{zhang2021direct}.

A much more flexible Targeted Information approach is \textbf{Shadow Tomography} (ST)~\cite{Aaronson_shadow}, a method designed to extract selected properties of a state, such as expectation values for selected observables, symmetries, entanglement properties and also fidelities, with as few measurement settings as possible. ST requires a mixed state to be tested and a set of $M$ observables designed to reflect the selected properties of interest~\cite{Aaronson_shadow}. More specifically, Aaronson~\cite{Aaronson_shadow} proposed an approach that counterintuitively allows extracting information about all $M$ selected observables without performing $M$ separate measurements. In particular, only $\widetilde{O}\left( \epsilon^{-4} \cdot \log^4 M \cdot \log d \right)$ copies of the state, where $d$ denotes the dimension of the system ($d = 2^n$ for a system with $n$ qubits) and $\epsilon$ the additive error, are sufficient to extract accurate predictions for all $M$ observables. Each observable is designed to test for multiple specific properties of the quantum state simultaneously, which are true or false for the tested mixed state depending on the measurement outcomes. The protocol achieves its efficiency through an adaptive measurement strategy that applies ''gentle measurements'' to minimize disturbance to the quantum state. ST maintains a set of quantum states consistent with all measurements performed so far and iteratively refines this set until sufficient precision is achieved for all target observables through a computationally expensive ''post-selection'' process. Further works extended ST to GST~\cite{helsen2023shadow} and QPT~\cite{kunjummen2023shadow}.

Inspired by these results, Huang et al.~\cite{Huang2020_cst} developed \textbf{Classical Shadows}, a practical implementation of the ST concept. A Classical Shadow is a classical approximate description of a quantum state, which enables the estimation of different observables with high probability using a small number of measurements. The core idea is that it is possible to efficiently store “snapshots” of the state, i.e., the measurement results for the selected measurement basis, for a specific measurement. A Classical Shadow is a set of such snapshots, containing sufficient information to predict linear functions of observables applied to the state.

To obtain the Classical Shadow of a quantum state $\rho$, the state is first projected probabilistically by applying a randomly selected unitary transformation $U$, such as a Clifford unitary, and then performing a measurement in the computational basis, resulting in the bitstring $b$. The randomness in the projection ensures that the different measurements capture different properties of the state. Afterwards, the snapshot is calculated as:
\begin{equation}
    \label{eq:snapshot}
    U^\dagger \ket{b}\bra{b}U,    
\end{equation}
and can be efficiently stored classically, representing one measurement outcome. This process is repeated to collect $N$ snapshots. The average of the $N$ snapshots can be seen as a measurement channel $\mathcal{M}$. Inverting this channel and applying $\mathcal{M}^{-1}$ to a snapshot, it is possible to obtain an estimate of $\rho$, denoted as $\hat{\rho}$. The Classical Shadow of $\rho$ is a set of $N$ such estimates obtained by applying $\mathcal{M}^{-1}$ to the $N$ collected snapshots:
\begin{equation}
    \label{eq:classical_shadow}
    S(\rho, N) = \left\{ \hat{\rho}_1, \hat{\rho}_2, \ldots, \hat{\rho}_N \right\}
\end{equation}

As the Classical Shadow is an approximation of $\rho$, it is possible to estimate any observable through the empirical mean:
\begin{equation}
    \braket{O} = \frac{1}{N}\sum_i \text{Tr}\hat{\rho}_i O.
\end{equation}

It must be noted that the process of inverting the measurement channel $\mathcal{M}$ is remarkably efficient for certain groups of measurements, such as Pauli and Clifford measurements. Furthermore, using Classical Shadows, it is possible to estimate the fidelity between any prepared state and any pure target state, with a number of measurements independent of system size \cite{Huang2020_cst}. Classical shadows were also extended to QPT~\cite{classical_shadow_qpt} and to broader classes of unitary ensembles~\cite{hu_cst}. Building on the original approach, enhanced protocols based on tensor networks have been introduced to further improve the efficiency of classical shadow construction~\cite{Akhtar2023scalableflexible_cst}.

\paragraph{Statistical and Machine Learning Tomography} 

\noindent\makebox[\columnwidth]{%
    \fbox{%
        \parbox{0.95\columnwidth}{%
            Statistical and Machine Learning Tomography techniques incorporate various statistical frameworks, including Maximum Likelihood Estimation (MLE), Bayesian inference and modern deep learning approaches, to generate a model of the underlying state or process. Using these frameworks, estimations of the characterized states or processes can be calculated with orders of magnitude fewer data points or iterative estimation steps in contrast to linear inversion based standard QST and QPT.
        }%
    }%
}
\\

The method to estimate a mathematical reconstruction for standard QPT and QST approaches was originally based on \textbf{Linear Inversion}. As the statistics of measurement results are a linear combination of the DM, the latter can be obtained from the former by solving a system of linear equations~\cite{ashburn1990experimentally, smithey1993measurement}. However, as highlighted by Hradil~\cite{hradil_mle_qst}, estimates based on inversion do not guarantee the positive definiteness of the reconstructed density or process matrix, leading to physically infeasible results. To address this issue, based on a core statistical concept, \textbf{MLE}, an improved reconstruction method was introduced, with an explicit and positive definite DM formula and a likelihood function for the observed measurements~\cite{hradil_mle_qst, measurement_of_qubits_james}. MLE makes it possible to incorporate additional information about the DM into the reconstruction process. Further MLE variants were introduced for tomographic reconstruction, such as an MLE algorithm with a tunable parameter controlling the trade-off between the convergence rate and the guarantee on the likelihood increase~\cite{vrehavcek2007diluted} and a scalable MLE algorithm for which all necessary operations can be implemented using matrix product states and operators~\cite{Baumgratz_2013_mle_qpt}.

Despite being intuitive and providing the possibility of imposing constraints, MLE is often inaccurate for QST and QPT, as it tends to estimate pure states even if the true state is mixed. This limitation led to the development of alternative estimators~\cite{Acharya2023} such as the particularly accurate \textbf{Projected Least Squares Estimator} (PLSE) for both QST~\cite{guctua2020fast} and QPT~\cite{surawy2022projected}, as well as estimators motivated by Bayesian statistics. The \textbf{Bayesian Mean Estimator} is not biased toward pure states, as it never yields zero eigenvalues and provides natural error bars for the accuracy of the estimate~\cite{blume2010optimal}. Efficient, numerically tractable \textbf{Bayesian point and region estimators} have also been proposed for QPT~\cite{Granade_2016_bayesian_qpt}. The advantages of Bayesian estimators come at the cost of finding a good prior distribution for the tomography~\cite{blume2010optimal}. An interesting proposal in this branch, \textbf{Fast Bayesian Tomography}~\cite{fast_bayesian_qpt}, uses prior information about probed gates, such as average fidelities obtained through randomized benchmarking, to improve tomographic quality and efficiency. This method allows for online learning: As more data becomes available, the estimates can be updated and improved in real time.

Methods from the realm of Machine Learning have also shown promising results for the estimation of quantum states and processes, taking partial measurement datasets as inputs and producing targeted reconstructions. For instance, Torlai et al.~\cite{Torlai2018} proposed the framework of \textbf{Restricted Boltzmann Machine QST} (RBM-QST), where a neural network architecture is based on the RBM model, where the input layer describes the physical qubits, and the hidden layer approximates the wave function's amplitude and phase, reconstructing a target state. Increasing the number of neurons improves the accuracy of the wavefunction approximation~\cite{Torlai2018}. Beyond the RBM model, it has been shown that any standard neural network can be adapted to perform QST and QPT~\cite{koutny_nn-tomography}, including Convolutional Neural Networks~\cite{Schmale2022_cnn}, Unsupervised Generative Models~\cite{Carrasquilla2019, torlai2023quantum}, Conditional Generative Adversarial Networks~\cite{Ahmed_cgans}, and Transformer-based Models \cite{Cha_2022_attention, yu2025transformer}, among others, sometimes collectively referred to as \textbf{Neural Network Tomography} (NN-QST or NN-QPT) methods.

Finally, a promising addition to ML-based tomography is the use of gradient descent optimization approaches for the estimation process. One such approach, \textbf{Gradient Descent QPT} (GD-QPT)~\cite{ahmed2023gradient}, uses gradient descent to learn Kraus operators for QPT, matching the performance of Compressed Sensing and PLSE approaches, while combining the benefits of both approaches: Similar to CS, GD-QPT requires a small number of random measurements, and similar to PLSE, GD-QPT can be scaled to larger systems (at least up to five qubits). Another very recent addition to this branch of tomography, \textbf{GD-QST}, makes use of gradient descent as a highly efficient DM reconstruction method, achieving high estimation fidelity for a full-rank seven-qubit QST~\cite{gaikwad2025gradient}. Impressively, this reconstruction result, based on a simulation, was achieved on a consumer-grade laptop in under three minutes~\cite{gaikwad2025gradient}. Both GD-QST and GD-QPT use manifolds, such as the Stiefel manifold~\cite{hu2020brief}, to constrain the optimization process, ensuring that the reconstructed Kraus operators or density matrices possess valid physical properties at each iteration, such as hermiticity, trace preservation, and positivity.

\clearpage
\begin{landscape}

\scriptsize
\setlength{\tabcolsep}{4pt}

\begin{longtable}{|p{2.5cm}|p{2.7cm}|p{3.0cm}|p{3.2cm}|p{3.2cm}|p{1.5cm}|}
\hline
\rowcolor[HTML]{EBEAFF}%
\textbf{Method} & \textbf{Constraint to Apply Method} & \textbf{Information Gained} & \textbf{Measurement Complexity (Required Observables)} & \textbf{Sample Complexity (Required state copies)} & \textbf{Sources} \\ \hline
\endfirsthead

\hline
\rowcolor[HTML]{EBEAFF} 
\textbf{Method} & \textbf{Constraint to Apply Method} & \textbf{Information Gained} & \textbf{Measurement Complexity (Required Observables)} & \textbf{Sample Complexity (Required state copies)} & \textbf{Sources} \\ \hline
\endhead

\hline
\multicolumn{6}{r}{\small Continued on next page...}\\
\endfoot

\hline
\endlastfoot

\hline
Full QST & None & Full density matrix & $\mathcal{O}(d^2)$ where $d=2^n$ & $\mathcal{O}(d^2/\epsilon^2)\log(d/\epsilon)$ & \cite{Flammia_2012_compressed_sensing_analysis, 10.1145/2897518.2897585} \\ \hline

Compressed Sensing QST & Low-rank density matrix (rank $r \ll d$) & Rank-$r$ approximation of the full density matrix & $\mathcal{O}(rd \log d)$ & $\mathcal{O}(r^2 d^2 \log d)$ & \cite{Flammia_2012_compressed_sensing_analysis, gross2010compressedSensing} \\ \hline

PI QST & None (PI-state for full tomography) & PI part of the density matrix & $\mathcal{O}(n^2)$ & $\mathcal{O}(n^{\mathcal{O}(1)})$ (Inferred) & \cite{toth_pi_tomography} \\ \hline

CS-QST in PI Subspace & Low-rank density matrix, PI state & PI part of density matrix & $\mathcal{O}(n^2)$ (Inferred) & $\mathcal{O}(n^{\mathcal{O}(1)})$ & \cite{schwemmer2014comparisonTomography} \\ \hline

MPS Tomography & Small bond dimension & MPS estimation of the state & $\mathcal{O}(n)$ (Inferred) & $\mathcal{O}(n^3)$ (Inferred) & \cite{Cramer2010, lanyon2017efficient} \\ \hline

Direct Fidelity Estimation & Known target process & Process fidelity & $\mathcal{O}(1)$ & $\mathcal{O}(d)$ & \cite{flammia2011direct} \\ \hline

Shadow QST & None & Expectation values of $M$ observables & $M$ (chosen parameter) & $\mathcal{O}(\varepsilon^{-4} \log^4 M \log d)$ & \cite{Aaronson_shadow} \\ \hline

Restricted Boltzmann Machine QST & Non-random correlations & RBM approximating wavefunction & Not stated & $\sim$100 samples/basis ($n=8$); 6400 ($n=20$) & \cite{Torlai2018} \\ \hline

Standard QPT & None & Full process matrix & $\mathcal{O}(d^4)$ & $\mathcal{O}(d^4/\epsilon^2)$ & \cite{Flammia_2012_compressed_sensing_analysis} \\ \hline

Compressed Sensing QPT (I) & Low Kraus rank $r$ & Rank-$r$ approximation of the process matrix & $\mathcal{O}(rd^2 \log d)$ & Not specified & \cite{Flammia_2012_compressed_sensing_analysis} \\ \hline

Compressed Sensing QPT (II) & Process matrix is $s$-sparse & $s$-sparse approximation of the process matrix & $\mathcal{O}(s \log d)$ & Not specified & \cite{compressive_qpt} \\ 
\hline

Standard GST & A gate set containing $K$ gates is provided & Full process matrix for each operation in the gate set. & $\mathcal{O}(K d^4)$ (Inferred) & Not specified & \cite{blume2013robust} \\ \hline

\hline

\end{longtable}

\captionof{table}{Measurement and sample complexities of selected QST and QPT variants.}
\label{tab:quantum_tomography_methods}

\end{landscape}
\clearpage

\noindent

\subsection{Software Focus}\label{software}
The Software Focus category includes benchmarking techniques specifically tailored to software and algorithm developers involved in the design and optimization of quantum software and algorithms, as well as supporting tools such as compilers and error correction protocols. However, this category is also relevant for hardware developers seeking to assess and compare their hardware as a holistic system, independent of specific applications. 

Often, these benchmarks address fundamental mathematical problems without a direct application in mind, solved by algorithms that may later integrate into end-user applications. This evaluation occurs at the circuit level, using multiple qubits and gates to assess overall device performance. These benchmarks prioritize theoretical performance and algorithmic optimization, as well as system-wide hardware assessment, which ultimately impacts end-user applications. This mid-level is particularly important in this regard, with error correction being one example where even small efficiency gains in its overhead can determine whether a relatively small quantum advantage given by an algorithm achieves practical advantage.
In the following, we introduce six major classes of software-focused benchmarks: volumetric, randomized, algorithm-based, dequantization, error correction, and compiler benchmarks.

\subsubsection{Volumetric Benchmarks}
\noindent\makebox[\columnwidth]{%
    \fbox{%
        \parbox{0.95\columnwidth}{%
        Volumetric benchmarks evaluate the performance by assessing the systems viability to execute quantum circuits as a function of qubit count (width) and consecutive gate count (depth). Adaptable in size, it measures scalability by defining operational success thresholds (e.g. fidelity or error rates) across varying problem sizes.
        }%
    }%
}
\\

A volumetric benchmark is characterized by a set of defining properties, including 1) a mapping, 2) constraint rules for execution, 3) a success metric for each circuit, 4) an overall success metric, and optionally 5) an experimental design~\cite{BlumeKohout2020volumetricframework}. First, a map from pairs of integers $(w, d)$ is defined, where $w$ corresponds to the circuit width (number of qubits), and $d$ represents the circuit depth (minimum number of steps to execute all gates in the circuit). This pair specifies an ensemble of test circuits $\mathcal{C}(w,d)$ used for benchmarking quantum devices. These test circuits may share a common structure and can be a single circuit $C$ or a list of circuits $\{C_1, ..., C_n\}$ that are all executed during testing. These benchmarks should include a rule specifying how the circuit should be compiled into the native gates of the quantum device, as well as a measure of success for each circuit, such as the heavy output probability~\cite{aaronson2016complexity, cross2019validating}, i.e., the fraction of experimental outputs falling into the set of ideally most-likely bitstrings. Additionally, an aggregated success measure depending on the specific benchmark's performance metric (e.g., again heavy output probability for quantum volume) is typically defined for the entire test circuit family, summarizing circuit performance. Some volumetric benchmarks also contain experimental execution guidelines that specify how circuits should be executed.

Volumetric benchmarks can accommodate different types of circuits, such as random, periodic, or application-based circuits. Examples of volumetric benchmarks include quantum volume~\cite{cross2019validating}, the cross entropy benchmark~\cite{boixo2018characterizing}, and Trotterized Hamiltonian simulation~\cite{lloyd1996universal}. Since volumetric benchmarks employ diverse circuit ensembles and evaluation methods, direct comparison between them is inherently challenging. The scalability of these circuit-level benchmarks strongly depends on their structure: While the first volumetric benchmarks, such as the earlier protocols for measuring quantum volume~\cite{moll2018quantum, cross2019validating}, face significant scalability challenges, newer approaches, such as mirror-circuits, offer a way to create scalable benchmarks. While versatile, the framework can be sensitive to \textit{SPAM} errors, making it useful for analyzing noise effects and overall system performance.

\begin{figure}[htbp]
\centering
\begin{tcolorbox}[title=BOX III, coltitle=white, colbacktitle=blue!75!black, sharp corners=south]
\textbf{Overview of Volumetric Benchmarks}

Volumetric Benchmarks provide a framework for evaluating QC performance across different hardware platforms by analyzing system capabilities as a function of executable circuit sizes. The circuits can follow a square format, where the width and depth of the circuit are identical as used by IBM's original quantum volume~\cite{cross2019validating}, but can also be rectangular, where the width and depth differ, as used in cross entropy benchmarking~\cite{boixo2018characterizing}. Each benchmark defines one or more criteria after which an executed circuit is deemed to have been successfully completed. These benchmarks are typically visualized in volumetric plots, mapping quantum system performance across different circuit sizes, depending on width and depth. We will now take a closer look at three volumetric benchmarking techniques below:
\vspace{0.5em} 
\begin{itemize}[nosep]
    \item Quantum Volume
    \item Mirror-circuit Benchmarking
    \item Cross Entropy Benchmarking
\end{itemize}
\end{tcolorbox}
\end{figure}

\paragraph{Quantum Volume}
Quantum volume, as defined by IBM in Ref.~\cite{cross2019validating}, assesses the performance of quantum computers by determining the largest random square circuits that can be reliably executed on its processor. The evaluation of whether a circuit has been successfully executed is defined differently from benchmark to benchmark and can follow different approaches (i.e. the observed error rate is below a certain threshold). It is denoted as $V_Q$ and defined via
\begin{equation}
\log_{2} V_Q = \argmax_m \min(m, d(m)).
\end{equation}

Here, $m$ is the number of qubits, and $d(m)$ represents the number of qubits in the largest square circuit for which it is possible to sample heavy outputs with a probability greater than $2/3$. Heavy outputs are the bit-strings whose ideal probabilities, according to the target quantum circuit, exceed the median of the ideal output distribution. 

Quantum volume is a circuit-level benchmark influenced by connectivity, gate parallelism, and error rates. It scales exponentially with the system size due to the need for classically simulating the circuits to compute the heavy outputs, as well as it is sensitive to noise and gate imperfections. Its standardized protocol~\cite{cross2019validating} allows for comparisons across diverse quantum architectures. However, quantum volume results are highly sensitive to factors such as the compiler, the compilation strategy, the native gate set, and the specific protocol employed. For example, an architecture with a larger native gate set may achieve higher practical performance, despite slightly lower per-gate fidelities, because fewer gates are required during compilation compared to architectures with more limited gate sets. Moreover, it was shown that experimentally measured quantum volumes frequently trail behind the values reported by the hardware manufacturers~\cite{pelofske2022quantum}, which once again emphasizes the importance of critically questioning benchmark results.

Quantum volume was first introduced by Bishop et al.~\cite{bishop2017quantum} as a performance metric for quantum computers, while Moll et al.~\cite{moll2018quantum} provided a formal definition and expansion on its applicability. Cross et al.~\cite{cross2019validating} later refined and standardized the concept, establishing it as a widely accepted benchmark for assessing the effective capabilities of quantum devices.

\paragraph{Mirror-circuit Benchmarks}
Mirror-circuit benchmarking evaluates quantum device performance by executing a quantum circuit followed by its inverse, with an intermediate layer of random Pauli gates. This construction ensures that, in the absence of noise, the output state corresponds to a known reference state, enabling efficient verification.

A mirror circuit $M(C)$ typically starts with the preparation of a random single-qubit state on each qubit. The chosen circuit $C$, which may include both single- and multi-qubit gates from the device’s native gate set, is then executed. This is followed by a layer of randomly selected single-qubit Pauli gates $Q$, after which the inverse circuit $\tilde{C}^{-1}$ is applied. In the absence of errors, the overall transformation  
\begin{equation}
\tilde{C}^{-1} Q C
\end{equation}
produces a deterministic and easily computable output bit-string for circuits of any size. Formally, this is expressed as  

\begin{equation}
\left| \psi_{\text{out}} \right\rangle = \Tilde{C}^{-1} Q C \left| \psi_{\text{in}} \right\rangle.
\end{equation}

Positioned at the circuit level, mirror circuits are sensitive to noise and errors while providing efficiently verifiable outputs. However, due to the circuit structure it is possible that systematic (coherent) errors in the implementation of $C$ cancel with the systematic errors in the implementation of $\Tilde{C}^{-1}$~\cite{proctor2022measuring}. The performance is quantified as the probability of observing the ideal output, which reflects the original circuit's accuracy. 

Proctor et al.~\cite{proctor2022measuring} introduced mirror circuits as a benchmarking method for the performance of quantum computers, which can also be used to determine the executable circuit sizes of a system. Later they extended this approach to estimate the infidelity of n-qubit circuit layers~\cite{proctor2022scalable}.

\paragraph{Cross Entropy Benchmarking}
Cross Entropy Benchmarking (XEB) assesses the performance of quantum computers by estimating the cross entropy between the actual and ideal, noise-free outcome distributions of executed circuits, therefore representing a circuit-level benchmark.

The experimental estimation takes random quantum circuits with $n$ qubits and $m$ layers of one- and two-qubit gates from an available universal gate set, samples their outputs, and compares their experimentally obtained output to ideal, noiseless circuit outputs. Therefore, it estimates the fidelity as
\begin{equation}
F_{\mathrm{XEB}} = 2^n \langle P(x_i) \rangle - 1,
\end{equation}
where $P(x_i)$ are probabilities from numerical, noise-free simulations, and $\langle \cdot \rangle$ denotes the average taken over all measured bit-strings $x_i$ obtained from the experiment. This normalization assumes that a completely noisy device outputs each of the $2^n$ possible bit-strings with equal probability $1/2^n$, corresponding to $F_{\mathrm{XEB}} = 0$, while $F_{\mathrm{XEB}} = 1$ corresponds to ideal circuit execution. This fidelity definition is a linear rescaling of the raw cross entropy~\cite{boixo2018characterizing}, mapping uniform output distributions to zero and perfect executions to one. 

The benchmark requires considerable classical resources, as the simulation of deep quantum circuits with many qubits scales exponentially. Multiple random circuit instances are typically averaged to reduce statistical fluctuations. Noise significantly impacts results, making XEB sensitive to errors while providing a rigorous test for quantum supremacy claims, including, for example, Refs.~\cite{arute2019quantum}, \cite{wu2021strong} or \cite{neill2018blueprint}.

\subsubsection{Randomized Benchmarks}

\noindent\makebox[\columnwidth]{%
    \fbox{%
        \parbox{0.95\columnwidth}{%
        Randomized Benchmarking (RB) is a protocol for estimating the average error rate of quantum gates, in which random operation sequences are applied and then inverted, and the resulting measurement results are fitted to an exponential decay curve.
        }%
    }%
}
\\

\begin{figure}[htbp]
\centering
\begin{tcolorbox}[title=BOX IV, coltitle=white, colbacktitle=blue!75!black, sharp corners=south]
\textbf{Overview of Randomized Benchmarking Techniques}

RB protocols offer, under particular limitations, a scalable, \textit{SPAM} robust, and hardware-agnostic benchmarking method to estimate average quantum gate fidelities~\cite{magesan2012characterizing}. Over time, a variety of RB techniques have emerged to address different experimental goals and hardware limitations. While standard RB characterizes the average performance of a gate set (typically Clifford), extensions adapt the protocol to isolate specific gate fidelities, measure crosstalk, detect leakage, or operate on non-Clifford gate sets. The diversity of RB variants reflects the need to probe increasingly nuanced error mechanisms in quantum systems, enabling a more complete and flexible assessment of hardware quality.
\vspace{0.5em} 
\begin{itemize}[nosep]
    \item Dihedral Randomized Benchmarking (DRB)
    \item Direct Randomized Benchmarking (Direct RB)
    \item Interleaved Randomized Benchmarking (IRB)
    \item Error-Specific Randomized Benchmarking
    \item Cycle Benchmarking (CB)
\end{itemize}
\end{tcolorbox}
\end{figure}

RB protocols, which differ in gate sets and targeted error types, generally consist of two phases: data collection and data processing~\cite{helsen2022general}. In the data-collection phase, randomly selected sequences of quantum gates, typically drawn from the Clifford group for efficient implementation, are applied to an initial state, usually $\ket{0}$, followed by an inversion operation intended to ideally return the system to its initial state. The final state is measured, and by averaging over multiple random sequences of quantum gates of varying lengths $m$ (where $m$ denotes the sequence length, e.g., the number of Clifford layers), the decay of the probability of obtaining the expected outcome is observed. In the data-processing phase, the set of survival probabilities $\{p_m\}_m$ is analyzed, where $p_m$ is the \emph{average probability} that the entire multi-qubit register is measured in the expected output state after an $m$-length sequence, averaged over many random realizations. This set is typically fitted to an exponential decay model, allowing for the extraction of average gate error rates:
\begin{equation}
    p_m \approx A + B f^m,
\end{equation}
where $A$ and $B$ are constants that encapsulate \textit{SPAM} errors, and $f$ is the decay parameter, which directly relates to the average fidelity of the gate set~\cite{helsen2022general}. Apart from these constant offsets, the survival probability is proportional to $f^m$.

However, the interpretation of the RB decay parameter as a direct measure of average gate fidelity has been critically re-evaluated. Proctor et al.~\cite{proctor2017randomized} show that due to gauge freedom in representing quantum operations, the RB decay is not uniquely related to fidelity, challenging the assumption that RB always yields a physically meaningful error rate. Depending on the proper representation, meaningful results can be generated. 
Conceptually, RB and mirror-circuit benchmarking differ in scope and methodology: RB averages over many randomized gate sequences to estimate the average error rate of a gate set, while being largely robust to \textit{SPAM} errors. In contrast, mirror-circuit benchmarks execute a specific circuit, followed by its inverse (interleaved with random Pauli gates), to directly measure the fidelity of that particular circuit’s execution. RB therefore characterizes average gate quality, whereas mirror-circuit benchmarking characterizes more the performance of a given circuit structure.

\paragraph{Dihedral and Direct Randomized Benchmarking}
Dihedral Randomized Benchmarking (DRB)~\cite{carignan2015characterizing, cross2016scalable} extends standard RB, focusing on the Dihedral group by incorporating non-Clifford operations, specifically adding phase gates, necessary for universal QC. DRB enables the assessment of coherent gate errors, such as systematic over- and under-rotations, and provides a refined fidelity metric relevant for determining error thresholds in fault-tolerant quantum computation.

Direct Randomized Benchmarking (Direct RB)~\cite{proctor2019direct}, in contrast to standard RB, operates on native gates without requiring decomposition into standard sets like Clifford gates and is therefore able to measure the gates' fidelity without decomposition errors. By tailoring random gate sequences to the target gate set, Direct RB efficiently characterizes native gates, which is especially beneficial for hardware platforms with unique native gates that lack an easy and efficient mapping to standard sets. Nevertheless, DRB is not fully scalable, as its SPAM routines demand for $n$ qubits $\mathcal{O}(n^2 / \log n)$ two-qubit gates.

\paragraph{Interleaved Randomized Benchmarking}
Interleaved Randomized Benchmarking (IRB)~\cite{magesan2012efficient} extends standard RB by providing a direct method for benchmarking the fidelity of individual quantum gates rather than an entire gate set. Instead of applying purely randomized sequences, IRB interleaves a specific target gate within a standard RB sequence, allowing its fidelity to be isolated and measured separately. The technique involves two benchmarking experiments: One using a standard RB sequence and another where the target gate is inserted at regular intervals.
The gate error of the target gate $G$ is estimated as follows:
\begin{equation}
    r_G^{\textnormal{est}} = \frac{(d-1)(1-p_{\bar G}/p)}{d},
\end{equation}
where $d = 2^n$ is the dimension of the system, $p$ is the depolarizing parameter from the first experiment, and $p_{\bar G}$ is the depolarizing parameter from the second experiment. 
For $r_G^{\textnormal{est}}$ it is necessary to be within the range $[r_G^{\textnormal{est}} - E, r_G^{\textnormal{est}} + E ]$, where 
\begin{equation}
    E = \min \left\{ \begin{array} {ll} \frac{(d-1) [|p-p_{\bar G}/p| + (1-p)]}{d} \\ \frac{2(d^2 -1)(1-p)}{pd^2} + \frac{4\sqrt{1-p}\sqrt{d^2-1}}{p}. \end{array} \right.
\end{equation}

This gate-level benchmarking approach is robust to \textit{SPAM} errors, as it averages the sequence fidelity over the survival probability. The resulting coefficients absorb \textit{SPAM} contributions, making the method a reliable tool for evaluating hardware-specific gate performance. However, unlike Direct RB, which evaluates native gate sets without decomposition, IRB assumes that errors in the interleaved gate are independent of the surrounding sequence, a condition that may not always hold. This phenomenon arises because unitary errors can coherently combine or cancel out with errors in the native gate set, potentially leading to slower decay of the interleaved sequence compared to the reference sequence, resulting in an unphysical negative error rate in IRB~\cite{hashim2024practicalintroductionbenchmarkingcharacterization}. Additionally, for IRB to provide meaningful results, the interleaved gate should ideally be part of the processor’s native gate set, ensuring an accurate assessment without introducing additional decomposition errors.

\paragraph{Error-Specific Randomized Benchmarking}
Beyond group-based RB extensions, Error-Specific Randomized Benchmarking has been developed to isolate and quantify specific errors in quantum hardware. Unlike standard RB, which provides a general error rate of the investigated gates, these techniques adapt gate sequences and/or analysis frameworks to enhance sensitivity to targeted error sources. This involves tailoring randomized sequences of gates to amplify the targeted error and analyzing the so-called survival probability, which is the overlap between the final noise-free simulated state and the actually measured state, through error-specific decay models. For example, Leakage Benchmarking targets errors in which quantum information escapes the computational subspace --i.e. the confined, finite-dimensional Hilbert space used for quantum computation-- into a larger, inaccessible subspace~\cite{chasseur2015complete, wallman2016robust, wood2018quantification, wu2024leakage}. Crosstalk Benchmarking quantifies unintended interactions between qubits, which can degrade performance during parallel gate execution~\cite{ketterer2023characterizing}. Unitary Error Benchmarking focuses on coherent, systematic gate calibration error, such as over-rotations or axis misalignment's~\cite{wallman2015estimating}. These specialized techniques provide fine-grained diagnostics that are particularly useful for optimizing performance at the lower levels of the hardware stack. However, they typically require careful circuit design and tailored gate sequences. Additionally, their analysis often depends on specific assumptions about the nature of the targeted errors, for example, that the modeled noise is Markovian and time-independent for leakage benchmarking~\cite{wu2024leakage}, making them more complex than standard RB.

\paragraph{Cycle Benchmarking}
Cycle Benchmarking (CB), developed by Erhard et al.~\cite{erhard2019characterizing}, evaluates quantum circuits by analyzing process fidelity across quantum gate cycles that are defined as a parallel set of gates acting on disjoint sets of qubits. CB quantifies how closely noisy gate operations approximate the ideal transformation, fitting results to error models to distinguish between coherent (i.e. unitary) and incoherent noise. Unlike standard RB, CB assesses multi-qubit gate cycles without gate set restrictions, enabling detailed error accumulation analysis. It is particularly suited for benchmarking deep circuits and multi-qubit systems, where interaction effects are critical. However, CB’s accuracy depends on selecting representative quantum gate cycles~\cite{calzona2024multi} and the validity of the assumed noise (models)~\cite{chen2023learnability}.

\subsubsection{Algorithm-based Benchmarks}
\noindent\makebox[\columnwidth]{%
    \fbox{%
        \parbox{0.95\columnwidth}{%
        Algorithm-based Benchmarks (ABBs) measure hardware performance based on quantum algorithms such as Shor's algorithm or quantum algorithmic subroutines like the Quantum Fourier Transformation (QFT). Key performance indicators include the scaling of success probability with respect to increasing problem size.
        }%
    }%
}
\\

Positioned as software-developer-oriented benchmarks, ABBs assess hardware capabilities by measuring performance metrics like gate fidelity, execution time, error rates, and scalability by executing specific quantum algorithms and subroutines. Typical benchmark subroutines include State Preparation~\cite{PhysRevLett.129.230504}, Block-Encoding~\cite{10012045}, Amplitude Amplification~\cite{brassard2002quantum}, and the QFT~\cite{coppersmith2002approximate}, which are critical building blocks for more complex quantum applications. Further, the algorithms of Shor and Grover have naturally been popular test cases \cite{365700,grover1996fast}. Often focusing on algorithms that can be scaled down enough for current hardware capabilities, other algorithms such as the Bernstein-Vazirani algorithm and hidden-shift algorithms have also been used to benchmark hardware performance~\cite{Wright2019}. For any given quantum algorithm, AABs yield error rates, resource requirements, and metrics related to algorithm success, such as the accuracy of Fourier coefficients in QFT or the amplification factor in Amplitude Amplification. These algorithm-specific success metrics are a key factor differentiating ABBs from randomized benchmarks, as small errors can have a significantly larger impact in highly structured circuits as opposed to randomly assembled ones~\cite{BlumeKohout2020volumetricframework}.

The impact of noise when scaling algorithms, such as with larger qubit registers and an increasing number of iterations in iterative algorithms, is a key aspect of these benchmarks. For example, a QFT benchmark could assess how well the system scales as the number of qubits increases, revealing potential bottlenecks in coherence time or gate fidelity. Similarly, Amplitude Amplification benchmarks may examine the impact of iterative processes on overall algorithmic accuracy and error accumulation.

This type of benchmark provides a direct assessment of the system's suitability for real-world applications by identifying its ability to perform key quantum operations reliably. Additionally, it highlights the impact of hardware-specific features like qubit connectivity, noise levels, and gate fidelity on algorithmic performance (see, e.g., Ref. \cite{mayer2024benchmarkinglogicalthreequbitquantum}). As reviewed by Georgopoulos et al.~\cite{georgopoulos2021quantumcomputerbenchmarkingquantum}, these insights can significantly support guiding the development of quantum hardware by linking algorithmic performance to practical application potential.

\subsubsection{Dequantization Benchmarks}
\noindent\makebox[\columnwidth]{%
    \fbox{%
        \parbox{0.95\columnwidth}{%
        Dequantization Benchmarks aim to compare quantum algorithms with their dequantized classical equivalents, which are derived from the original quantum algorithm and generally exhibit similar complexity under specific assumptions. 
        }%
    }%
}
\\

This benchmark category is mainly of interest to software developers, end-users, and domain experts, targeting problems for which quantum algorithms initially demonstrated theoretical advantages. Dequantization benchmarks yield quantum vs. classical comparisons of, e.g., runtime, space, error scaling, and fidelity of results for given algorithms applied to specific problem instances like matrices for solving linear systems of equations or machine learning datasets for binary classification.

De-quantized algorithms often rely on problem instance-specific assumptions like sparsity, low-rank matrices, or approximate sampling distributions (see e.g., Ref.~\cite{PhysRevLett.127.060503}). The benchmark evaluates the feasibility of these assumptions in practical settings and their impact on the algorithm's scalability. For example, quantum algorithms for linear algebra (e.g., the quantum Principal Component Analysis~\cite{Lloyd2014a}) are compared to their classical counterparts derived through dequantization, such as the techniques proposed by Tang~\cite{PhysRevLett.127.060503}, which exploit sampling-based approaches to achieve comparable complexities under specific data conditions. Note, however, that the classical data access models should be chosen based on envisioned applications, as they have a significant potential to unintentionally be (exponentially) more powerful than the quantum data access models~\cite{cotler2021revisitingdequantizationquantumadvantage} which can lead to pseudo-dequantizations, which solve a different problem than the original quantum algorithm of interest.

This benchmark category also considers hardware-independent factors like classical preprocessing and hardware-specific factors like coherence times and qubit connectivity for the quantum version. It quantifies the crossover point where the quantum algorithm starts to outperform its classical counterpart or vice versa, providing critical insights for technology readiness assessments. By doing so, they evaluate whether the quantum advantage claimed for a specific problem holds under real-world conditions. 

Key literature includes Tang on classical sampling techniques inspired by quantum algorithms~\cite{10.1145/3313276.3316310}, Aaronson on practical limitations of quantum speedups~\cite{Aaronson2015}, and Chia et al.~\cite{10.1145/3549524} generalizing initial dequantization techniques. Recent experimental validations, such as those on NISQ devices and classical hardware, showcase the practical relevance of this benchmark~\cite{PhysRevLett.131.100803}.

\subsubsection{Error Correction Benchmarks}
\noindent\makebox[\columnwidth]{%
    \fbox{%
        \parbox{0.95\columnwidth}{%
        Error correction based benchmarks are used to evaluate the efficiency of Quantum Error Correction (QEC) codes as well as error-corrected quantum processing units in a holistic manner. 
        }%
    }%
}
\\

Error correction is crucial for scalable QC by detecting and correcting errors caused by imperfections of the hardware implementation through using multiple physical qubits to implement logical qubits with significantly lower error rates. Popular QEC protocols include Steane codes~\cite{steane1996error}, surface codes~\cite{kitaev2003fault}, and color codes~\cite{bombin2006topological}, all of which differ in their induced quantum circuit overhead and corresponding error rates. These must not be confused with error mitigation techniques such as dynamical decoupling~\cite{viola1998dynamical}, zero-noise extrapolation~\cite{temme2017error, giurgica2020digital}, and probabilistic error cancellation~\cite{van2023probabilistic}, which merely reduce the effect of errors without an explicit in-circuit correction protocol and in turn require their own specialized benchmarking techniques~\cite{Bultrini2023unifying}.

A simple example of an error correction algorithm is the 3-qubit repetition code, which protects a single qubit against a bitflip error (i.e., the unwanted application of a Pauli X-gate)~\cite{PhysRevA.52.R2493}. There, we can observe the three steps of a quantum error correction code: (1) encoding followed by the possible error, (2) decoding, and (3) correction. This approach notably requires mid-circuit measurements and a classical computer to decide on the operations necessary for the error correction based on the syndrome measurement, which can be a significant bottleneck depending on the hardware implementation~\cite{Bausch2024}. Notably, first experiments have already shown that existing quantum computers are already capable of applying error correction that outperforms non-error-corrected implementations~\cite{Acharya2025}.

QEC benchmarks are targeted towards the evaluation of quantum computers at the compilation level as well as holistically, while focusing on the effectiveness and resource demands of error-handling strategies~\cite{Acharya2023}. Inputs typically include algorithmic quantum circuits or quantum states and the specifications of error correction codes, while outputs are performance indicators that can be structured into eight categories according to
Chatterjee and Ghosh~\cite{10821356}: (1) Qubit Overhead, i.e., the number of  physical qubits utilized to encode a logical qubit, (2) Error Threshold, i.e., the maximum physical error rate that still allows for applying the error correction code successfully (3) Error Protection, i.e., which types of errors the error correction code protects against, (4) Decoding, i.e., the availability and efficiency of required decoding algorithms, (5) Transversal Gates, i.e., the availability of gates that allow for operations on encoded qubits without decoding them, (6) Scalability, i.e., the ability of the error correction code to stay effective as the system size increases, (7) Realization, i.e., the versatility of error correction codes wrt. the underlying qubit technology used (trapped-ions, neutral atoms, etc.), and (8) Complexity, i.e., the structural and operational complexity of its encoding, detection, and correction steps. Depending on the desired output metric, comparisons with classical, noise-free circuit simulations or already discussed benchmarking techniques such as quantum tomography or randomized benchmarking can be applied using the implementation of the error-corrected quantum circuit as input~\cite{Acharya2025,wootton2020benchmarking}.

\subsubsection{Compiler Benchmarks}
\noindent\makebox[\columnwidth]{%
    \fbox{%
        \parbox{0.95\columnwidth}{%
        Quantum compiler benchmarking evaluates the efficiency of quantum compilers and circuit optimization techniques in transforming high-level quantum algorithms into low-level hardware-executable instructions. It assesses key performance metrics such as circuit depth, gate count reduction, compiler runtime, and hardware adaptability to quantify optimization effectiveness and scalability.
        }%
    }%
}
\\

Quantum compilers play a critical role in adapting abstract quantum algorithms to concrete hardware constraints. They manage qubit mapping, routing, and gate decomposition into native instructions, all while minimizing circuit depth and error accumulation. Because hardware-specific constraints—such as gate sets, qubit connectivity, and noise characteristics—vary widely across platforms, compiler benchmarking must be multifaceted. No single metric suffices; instead, comprehensive assessments incorporate indicators like two-qubit gate counts, total circuit depth, fidelity loss, and hardware-specific noise sensitivity.

\begin{figure}[htbp]
\centering
\begin{tcolorbox}[title=BOX V, coltitle=white, colbacktitle=blue!75!black, sharp corners=south]
\textbf{Overview of Compiler Benchmarking Techniques}

Benchmarking quantum compilers helps assess how well compilation workflows preserve algorithmic intent while adapting to physical constraints. Beyond optimizing gate count or circuit depth, compiler performance includes resilience to noise, efficiency of qubit routing, and the ability to scale with problem size. Evaluating these aspects enables comparison between compilation frameworks and guides the development of hardware-aware compilation strategies. Examples of benchmarks targeting these aspects include:

\vspace{0.5em}
\begin{itemize}[nosep]
    \item Quantum Layout Synthesis
    \item Quantum Circuit Unoptimization
\end{itemize}
\end{tcolorbox}
\end{figure}

\paragraph{Quantum Layout Synthesis}
The NP-complete problem of quantum layout synthesis~\cite{tan2020optimal} is about transforming quantum programs to satisfy hardware constraints. It involves translating quantum gates into the hardware's native gates set, mapping logical qubits to physical qubits, and scheduling the execution of each gate. A key aspect of layout synthesis is inserting SWAP gates to ensure two-qubit operations align with hardware connectivity.
Layout synthesis takes a quantum program and a coupling graph $(P,E)$, where $P$ are physical qubits and $E$ their connections. The output includes spacetime coordinates $(t_l,x_l)$ defining execution order, inserted SWAP gates, and a final mapping $\pi :Q \to P$ of logical to physical qubits. Since two-qubit gates must align with existing edges while avoiding collisions, gate cancellation and optimization have to precede layout synthesis. The main objectives are minimizing circuit depth, SWAP cost, and gate count while maximizing fidelity. The total number of variables can be reduced, depending on the formulation, to $\mathcal{O}(NT)$, where $N$ is the number of physical qubits and $T$ the upper time bound~\cite{tan2020optimal}.
Noteworthy here is the Arline Benchmark~\cite{kharkov2022arline}, a software framework for automating quantum compiler benchmarking, particularly for NISQ devices. It evaluates circuit optimization subroutines using metrics such as two-qubit gate count, circuit depth, runtime, and a hardware-dependent cost function that integrates depth, gate fidelity, and penalties for deep circuits. Compilation efficiency is assessed via the ratio of initial-to-final cost, with additional metrics including average log-fidelity and circuit compression factors. Arline operates from a configuration file specifying hardware, compilers, and circuits, producing detailed reports with figures, charts, and QASM outputs at each compilation stage. Benchmarks can be performed on both random and structured circuits, with performance measured by the proximity between ideal and measured output distributions.

\paragraph{Quantum Circuit Unoptimization}
Quantum Circuit Unoptimization, introduced by Mori et al.~\cite{mori2023quantum}, reverses circuit optimization by intentionally adding redundancies while maintaining functionality. It introduces gate insertions, swaps, decompositions, and synthesis steps to generate circuits that challenge compiler optimization. Unoptimization provides a controlled method to, for example, evaluate compiler effectiveness in reducing redundant operations, facilitating direct comparisons between compilation frameworks and assessing their impact on final circuit fidelity.
The unoptimization process generates a redundant circuit $V$ from the original circuit $U$, both expressed in the $\{U3, CX\}$ gate set, allowing for depth measurements before and after compilation. Two key metrics assess compiler efficiency: the unoptimized ratio $t_{\textnormal{unopt}}=d_{\textnormal{unopt}}/d_{\textnormal{original}}$, which quantifies redundancy in $V$, and the optimized ratio $r_{\textnormal{opt}}=d_{\textnormal{opt}}/d_{\textnormal{original}}$, indicating the compilers ability to reduce circuit depth. Thereby, $d_{\textnormal{original}}$ is the depth of $U$, $d_{\textnormal{unopt}}$ is the depth of $V$, and $d_{\textnormal{opt}}$ the depth of the optimized circuit. A more efficient compiler reduces the gap between $r_{\textnormal{opt}}$ and $1$, with the reduction from $t_{\textnormal{unopt}}$ to $r_{\textnormal{opt}}$ serving as a measure of the compiler's optimization effectiveness. In certain cases, the depth of the resulting circuit can even be lower than $d_{\textnormal{original}}$, if the original circuit was not minimal to begin with.

\subsection{Application Focus}\label{application}
Application-focused benchmarks are designed by and for quantum software engineers as well as domain specialists developing QC applications for real-world problems. These benchmarks evaluate the full QC stack —hardware, software, and applications— to identify optimal component integration for practical use and end-to-end efficiency.
The focus is on how components interact and collectively impact system-wide performance, particularly in practical QC workflows~\cite{rohe2025problem}. Given current NISQ-era constraints, these benchmarks often rely on small-scale demonstrators rather than full-scale applications. What they have in common is that they are based on real-world use cases, where performance is evaluated in the context of practically motivated problem instances, giving an indication of the actual capability of QC. 
In the following, we present four prominent classes of application-focused benchmarks: ground-state energy calculations, simulation of (quantum) physical processes, quantum machine learning (QML) benchmarks, and quantum optimization benchmarks.

\subsubsection{Ground State Energy Calculations} 

\noindent\makebox[\columnwidth]{%
    \fbox{%
        \parbox{0.95\columnwidth}{%
            Ground-state energy benchmarks evaluate a quantum computer’s ability to approximate the smallest eigenvalue or the associated eigenstate of a system’s Hamiltonian, assessing accuracy, convergence, and resource efficiency relative to classical methods and/or experimental data.
        }%
    }%
}
\\

Ground-state energy benchmarks assess a quantum computer's precision in determining the smallest eigenvalue (and/or the corresponding eigenstate) of a system's Hamiltonian, a critical factor in quantum chemistry for predicting molecular properties and behaviors. The ground state defines the most stable molecular configuration, influencing molecular geometry and reactivity. Moreover, the energy gap between the ground and excited states dictates reaction kinetics and activation energies, as described by transition state theory~\cite{barkoutsos2018quantum, cao2019quantum}. This computational problem is notoriously hard. The Hartree–Fock method—widely used as a classical baseline—is NP-complete, and more general ground-state energy problems are even QMA-complete. Classical methods such as Hartree–Fock and post-Hartree–Fock provide approximate solutions, often recovering around 99\% of the total electronic energy, but they fall short in strongly correlated systems~\cite{whitfield2013computational}. Quantum approaches are thus especially of high interest when retrieving exact solutions is crucial.

Typically, these benchmarks use the Variational Quantum Eigensolver (VQE)~\cite{peruzzo2014variational} as the primary quantum algorithm to iteratively minimize the expected energy of a trial quantum state, or ansatz, defined by variational parameters. Through an optimization loop involving both quantum subroutines and classical optimization algorithms, the VQE approximates the ground state energy of the underlying system. The quantum subroutine evaluates the system energy, while the classical optimizer adjusts the variational parameters. 

In addition to the VQE approach, adiabatic quantum algorithms, particularly adiabatic quantum computing (AQC), are also employed for ground-state benchmarks. AQC evolves a quantum system adiabatically from an easy-to-prepare initial state to the typically complex-to-prepare ground state of a target Hamiltonian~\cite{doi:10.1126/science.1113479,babbush2014adiabatic}. Although AQC has successfully determined ground states for simple molecules, like methylene ($CH_2$), for which, e.g., Ref.~\cite{10.1063/1.3503767} showed how to prepare both ground and excited states efficiently, there remains no general proof of polynomial-time solvability for arbitrary chemical systems~\cite{Lee2023}. Nonetheless, recent theoretical work provides some runtime bounds for adiabatic algorithms, heavily dependent on the spectral gap between the ground and first excited states~\cite{PhysRevResearch.5.033175,PRXQuantum.3.040327}. Particularly in chemical systems, the typically large spectral gap of real-world molecules raises hopes for computational efficiency, even though general theoretic guarantees for the entire evolution period remain unproven. Furthermore, evidence suggests adiabatic methods address computational complexity classes inaccessible to classical algorithms, indicating potential quantum advantages~\cite{10.1063/1.3598408}. Notably, the distantly related problem of computing the ground state energy, given the ground state can be solved exponentially faster than possible with known classical algorithms using Quantum Phase Estimation~\cite{PhysRevLett.83.5162}. 

The benchmarks measure quantum hardware and algorithm performance by comparing computed energies to ``chemical accuracy'' -- the threshold of deviation acceptable in chemical simulations~\cite{mccaskey2019quantum}. Furthermore, benchmarks consider the computational speed and efficiency of the process. These benchmarks can be based on various molecular and condensed matter systems and are suitable cross-platform performance analyzers~\cite{mccaskey2019quantum, lubinski2023application}. This approach evaluates the entire hardware-software stack with all its components, including error mitigation strategies to reduce hardware-induced noise.

In 2018, Hempel et al.~\cite{hempel2018quantum} implemented a trapped-ion quantum simulator to compute the ground state energies of hydrogen and lithium hydride molecules via the VQE, demonstrating different encoding methods and noise mitigation strategies. While not the primary focus of their study, this work contributed to early cross-platform application benchmarks for multiqubit quantum systems in quantum chemistry. Expanding on this, McCaskey et al.~\cite{mccaskey2019quantum} developed a benchmark for NISQ devices, computing ground-state energies of alkali metal hydrides via VQE while employing error mitigation techniques like reduced density matrix purification to improve accuracy. Using IBM and Rigetti quantum processors, their benchmark incorporated chemical accuracy as a performance metric and introduced a scalable approach for cross-platform comparisons. This benchmark enables the evaluation of both hardware and algorithmic progress while assessing the potential for quantum advantage over classical methods. In a related direction, Dallaire-Demers et al.~\cite{dallaire2020application} introduced a benchmark based on the one-dimensional Fermi-Hubbard model to evaluate NISQ devices' capability in simulating fermionic systems via ground-state calculations with the VQE. They introduced the effective fermionic length metric, quantifying a quantum device's ability to approximate ground-state energies across increasing chain lengths. A key advantage of this method is that one-dimensional Hubbard models are exactly solvable via the Bethe ansatz~\cite{bethe1931theorie}, offering a precise classical benchmark. The exact solvability of these models enables scalable benchmarking. Gard and Meier~\cite{gard2022classically} later applied a similar benchmarking approach using the one-dimensional Hubbard model. Further advancing this area, Yeter-Aydeniz et al.~\cite{yeter2021benchmarking} introduced a suite of alkali hydride molecules as benchmark systems for QC solutions.

\subsubsection{Simulating (Quantum) Physical Processes}
\noindent\makebox[\columnwidth]{%
    \fbox{%
        \parbox{0.95\columnwidth}{%
            Benchmarks that employ quantum algorithms to simulate (quantum) physical processes, ranging from statistical mechanics to molecular dynamics to evaluate key performance indicators like runtime and accuracy of a quantum system. 
        }%
    }%
}
\\

Since Feynman's proposal of using quantum computers for the simulation of quantum physical processes in 1982~\cite{Feynman1982}, which was ultimately proven to be correct in 1996 by Lloyd~\cite{lloyd1996universal}, much progress on efficient quantum algorithms for the simulation of quantum physical processes has been made~\cite{10.1145/780542.780546,10.5555/2481569.2481570,10.5555/2231036.2231040,Lloyd2014a,Berry15,PhysRevLett.118.010501,Low2019hamiltonian,PhysRevA.110.012612}. These algorithms enable the simulation of the unitary time evolution $e^{-i \hat{H} t / \hbar}$ guided by a given Hamiltonian, e.g., as a linear combination of unitaries $H=\sum_{j=1}^d \alpha_j U_j$, in time $\mathcal{O}(\log\left({1/\varepsilon}\right) + t\sum_{j=1}^d|\alpha_j|)$ using $\mathcal{O}(\log\left({d}\right))$ ancillary qubits, up to an error $\varepsilon$ in the spectral norm~\cite{Low2019hamiltonian}. Practical applications include, e.g., the time evolution of electronic systems~\cite{PhysRevX.13.041041}, and have been shown to require exponentially fewer computational resources than conventional classical methods.

Beyond the simulation of purely quantum mechanical processes, it has been shown that quantum algorithms can also speed up the simulation of physical systems evolving under the laws of classical mechanics by an up-to-exponential factor~\cite{PhysRevX.13.041041}. In general, many of these problems can be formulated as differential equations, which can be solved up to exponentially faster in many (potentially) relevant cases~\cite{Cao_2013,PhysRevA.99.012323,Linden2022,PhysRevX.13.041041,stein2024exponentialquantumspeedupsimulationbased} using a quantum linear system solver~\cite{HHL,Ambainis12,Berry15,Childs17,gilyen2018quantum,SSO19,Lin2020optimalpolynomial} and possibly linearization~\cite{leyton2008quantumalgorithmsolvenonlinear,lloyd2020quantumalgorithmnonlineardifferential,doi:10.1073/pnas.2026805118}.

Benchmarks for problems focusing on Hamiltonian simulation are still severely limited by current quantum hardware performance, as the employed algorithms usually require fault-tolerant QPUs, such that only very small problem instances can be investigated in the NISQ era. Nevertheless, first benchmark frameworks and metrics have been proposed in literature, which focus on quantum process fidelity in addition to circuit dimension, while predominantly relying on numerical simulations as baselines~\cite{Dong2022,granet2025appqsimapplicationorientedbenchmarkshamiltonian}.
After initial proofs of concept for the application of quantum linear system solvers on quantum hardware~\cite{cai2013experimental,zheng2017solving}, Yalovetzky et al.~\cite{Yalovetzky2024} recently benchmarked a specific variant of a quantum linear system solver on a trapped-ion quantum hardware, investigating the required circuit dimensions as well as the result fidelity through comparisons with noiseless numerical simulations. Collectively, these benchmarking studies provide essential insights into the practical capabilities and limitations of near-term quantum computers for simulating physical processes, informing future algorithmic and hardware developments.

\subsubsection{Quantum Machine Learning Benchmarks} 
\noindent\makebox[\columnwidth]{%
    \fbox{%
        \parbox{0.95\columnwidth}{%
            Quantum machine learning (QML) benchmarks evaluate QC systems by applying quantum machine learning algorithms to canonical learning tasks, such as classification, clustering, or generative modelling, thereby assessing the system’s ability to perform data-driven optimization across the computational stack.
        }%
    }%
}
\\

Machine learning spans supervised, unsupervised,  and reinforcement learning, each of which has quantum counterparts using quantum circuits and hybrid models for benchmarking~\cite{dawid2022modern}. The general goal of machine learning is to develop models that can learn patterns or make decisions from data, often by optimizing a set of parameters to minimize (or maximize) an objective function, enabling the model to perform well on new, unseen inputs. QML offers potential advantages over classical machine learning in expressivity~\cite{abbas2021power}, trainability~\cite{abbas2021power}, and runtime complexity~\cite{lloyd2013quantum, liu2021rigorous} -- but these advantages are problem-specific and currently constrained by hardware and data input limitations.

Its ability and efficiency, especially but not exclusively measured in terms of loss or accuracy, can be used to benchmark and compare algorithms, hardware components, or even the entire stack. Well-defined loss functions and labeled datasets enhance this benchmark's attractiveness. However, due to the broad design space of QML benchmarks, their neutrality must be critically reassessed on a case-by-case basis~\cite{bowles2024better}. Thereby, the selected baseline and the selected problem instances play a decisive role. 
These benchmarks evaluate the entire QC stack, requiring only task-specific data as input, while accuracy-based metrics define performance. Although current QML algorithms may scale poorly, benchmark scalability remains viable for larger systems. 

Since the rise of QML~\cite{biamonte2017quantum}, QML algorithms have been used increasingly to benchmark quantum hardware and full-stack system performance. Benedetti et al.~\cite{benedetti2019generative} introduced the qBAS score, a task-specific performance metric designed to evaluate different hybrid quantum-classical systems. The metric is based on the generative modelling performance on the Bars and Stripes (BAS) dataset, a canonical synthetic dataset that is classically easy to generate, visualize, and validate for sizes up to hundreds of qubits. The task poses a significant challenge for shallow quantum circuits due to the high requirement of entanglement, making it sensitive to device recalibrations and environmental noise – which in turn is good for benchmarking. Benedetti et al. evaluate the qBAS score experimentally using an ion-trap quantum device and their proposed data-driven quantum circuit learning (DDQCL) algorithm.
Similar work has been conducted by Hamilton et al.~\cite{hamilton2019generative}, who also employ generative modelling on the BAS dataset using the DDQCL approach to benchmark small quantum systems implemented with superconducting qubits. The performance is evaluated using the Kullback-Leibler divergence and two variations of the F1 score, which measure the similarity between the target distribution and the sampled distribution. A follow-up study expanded this work by comparing multiple QPUs with an emphasis on the effectiveness of error mitigation techniques~\cite{hamilton2020error}.
Bowles et al.~\cite{bowles2024better} highlight the significant influence of experimental design choices on benchmarking outcomes in QML, explaining how aspects like dataset selection, model tuning, and evaluation metrics can be exploited to portray quantum models more favorably than warranted. They discuss issues such as selective dataset use, reporting only the best-performing models, and unbalanced classical baselines as common sources of bias. To address these concerns, the authors introduce a dedicated benchmarking framework, which systematically tests a wide range of quantum models across multiple datasets. This framework aims to improve scientific rigor by enforcing reproducibility, reducing positivity bias, and enabling a structured comparison of design choices across model families. QML-based benchmarks can be found as a separate category in many benchmark suites~\cite{kiwit2023application}, as specific training and test datasets have been created for quantum machine learning applications~\cite{schatzki2021entangled, perrier2022qdataset}.

\subsubsection{Quantum Optimization Benchmarks} 
\noindent\makebox[\columnwidth]{
    \fbox{
        \parbox{0.95\columnwidth}{
            Quantum Optimization Benchmarks assess the quantum system performance by solving well-defined (combinatorial) optimization problems using quantum algorithms such as the Quantum Approximate Optimization Algorithm (QAOA).
        }
    }
}
\\

Quantum optimization benchmarks evaluate the performance of quantum computers by solving optimization problems and comparing results against reference points, such as classical heuristics, optimal brute-force solutions for small instances, or alternative quantum approaches~\cite{bucher2024towards}. Performance can be assessed in terms of solution quality, runtime, or other computational resource requirements. These benchmarks typically target NP-hard problems, where classical algorithms are inefficient, making them promising candidates to achieve quantum advantage. The motivation is to evaluate how well quantum algorithms can approximate or solve such problems, using metrics that reflect both solution quality and computational efficiency. Various optimization problems, such as the maximum cut (MaxCut) problem, the traveling salesman problem, or satisfiability problems, are frequently employed as benchmark tasks. Based on these problem classes, various problem instances are generated and solved. The performance is then measured based on solution quality, quantified through multiple metrics. The most well-known metric is the approximation ratio, the ratio between the quantum result and the optimal or best-known solution, although other metrics and quality criteria are also conceivable (e.g. the mean of the sampled energies or the $\alpha$-quantile of the sampled energies~\cite{barkoutsos2020improving}). Furthermore, time-to-solution~\cite{martiel2021benchmarking} or scaling behavior with problem size~\cite{sankar2024benchmarking} are additionally considered to reflect computational practicality.

The MaxCut problem is one of the most widely used problems for benchmarking here. It is defined on a graph consisting of nodes and edges, which may optionally include weights. The objective of this optimization problem is to partition the nodes into two disjoint groups such that the number of edges between the groups (if so, the sum of their weights) is maximized. This problem is NP-hard and has an efficient classical approximation algorithm, the Goemans-Williamson algorithm, which guarantees an approximation ratio of at least $0.8786$ times the optimal solution for classical computation~\cite{goemans1995improved}. Due to its natural formulation as a quadratic unconstrained binary optimization (QUBO) problem, the MaxCut problem maps directly to the Ising model, making it particularly well-suited for quantum algorithms such as the QAOA or quantum annealing. However, from an industrial perspective, MaxCut is of limited direct relevance, as real-world optimization problems often involve complex constraints and more intricate mathematical formulations that require richer encoding schemes~\cite{ruan2023quantum}.

Unlike QML benchmarks, which focus on iterative learning and generalization, quantum optimization benchmarks emphasize solving individual problem instances. To meaningfully assess computational performance, it is crucial to select problem instances that are intrinsically hard —often those near computational phase transitions— where the structure of the solution space changes abruptly, and solvers experience maximal difficulty~\cite{monasson1999determining, zhang2022quantum}. These benchmarks enable evaluation across the full quantum stack, from problem encoding and compilation to execution on quantum hardware. Inputs typically include problem-specific data such as graph structures, cost functions, or constraint matrices. While certain benchmark formulations can theoretically scale to large qubit numbers, meaningful analysis of scaling behavior remains challenging in practice due to current hardware limitations.

Optimization problems, especially combinatorial ones, have long been a natural choice for benchmarking QC capabilities~\cite{crooks2018performance, streif2019comparison}. In the early phases of quantum software and algorithm development, optimization-based benchmarks primarily targeted the evaluation of algorithmic performance~\cite{willsch2020benchmarking}. A prominent full-stack benchmark using an optimization problem is the \textbf{Atos Q-score\textsuperscript{TM}}, which evaluates quantum system performance through the MaxCut problem solved with QAOA~\cite{martiel2021benchmarking}. The Q-score\textsuperscript{TM} is defined by the maximum number of qubits that can be effectively utilized to solve problem instances generated from Erd\H{o}s-R\'enyi graphs. These graphs are constructed with an edge connectivity of $p = 0.5$, and the number of nodes corresponds to the number of qubits involved. For each graph size $n$, $100$ random problem instances are generated and solved using the QAOA. The energy measurements from these solutions are used to compute a score, $\beta$, defined as the normalized difference between the average measured cut energies $C(n)$ and the expected value from random guessing $\frac{n^2}{8}$. 
Here, $\lambda$ is an empirically determined constant characterizing the $n^{3/2}$ scaling of the optimal MaxCut for such graphs, serving to normalize the metric so that $\beta(n)$ measures the quantum algorithm’s success as a fixed fraction of the maximal possible improvement over random guessing when problem size increases:

\begin{equation}
\beta(n) = \frac{C(n) - \frac{n^2}{8}}{\lambda n^{3/2}}.
\end{equation}

The benchmark is considered passed if $\beta$ exceeds a predefined threshold $\beta^*$, typically set to 0.2. This threshold, ranging from $0$ (indicating random performance) to $1$ (representing an exact solver), defines the minimal solution quality required to pass. While the choice of $\beta^*$ is somewhat arbitrary and could, in principle, be adjusted to influence the benchmark's difficulty, it is fixed in practice to ensure fair comparisons across systems. By iteratively testing increasing numbers of qubits, the Q-score\textsuperscript{TM} identifies the largest problem size that the hardware-software combination can adequately solve according to the set quality standard. A significant advantage of this benchmarking approach is its inherent scalability, as it enables performance evaluation across different system sizes while mitigating distortions from purely classical scaling behaviors.

The \textbf{QPack} benchmark~\cite{mesman2021qpack} extends optimization-based benchmarking by evaluating quantum systems across multiple combinatorial problems and performance dimensions, such as runtime, accuracy, and scalability. Building on this, QPack Scores~\cite{donkers2022qpack} introduce a structured scoring system with distinct sub-scores, providing a holistic assessment of QC platforms.

\section{Interconnectedness  of Benchmarks}\label{interconnectedness}
Benchmarks at each level influence and constrain the others in a top-down and bottom-up fashion. At the lowest level, hardware metrics set the fundamental performance envelope for the entire system. For example, qubit count and connectivity limit which algorithms can be run and with what overhead: A sparse connectivity necessitates additional SWAP operations during circuit mapping, increasing circuit depth and error accumulation for applications. Likewise, (too) limited gate fidelity and decoherence times mean that as algorithms require more operations or qubits, their success probabilities drop exponentially if errors compound. In other words, fidelity metrics obtained from low-level benchmarks directly bound the feasible circuit depth and scale for higher-level algorithms. This fundamental dependency is why improving hardware error rates (as revealed by tomography or RB) immediately expands the possibilities at the application level. Indeed, certain low-level benchmarks are explicitly linked to high-level viability: For instance, RB variants can yield error estimates relevant to fault-tolerance thresholds, indicating whether and when the hardware is ready for error-corrected algorithms. 

Mid-level benchmarks serve as a critical bridge between isolated hardware metrics and end-user outcomes. They often translate hardware performance into operational capabilities. Volumetric benchmarks like quantum volume characterize a device’s capability to run increasingly complex circuits by sweeping over qubit count and circuit depth, defining an ``operational success'' threshold (e.g. a minimum fidelity) for each size. Such metrics effectively build up on low-level error rates and output a single figure of merit (e.g. Quantum Volume). Similarly, ABBs at this mid-level execute standard algorithmic subroutines (QFTs, amplitude estimation, etc.) to probe the system’s integrated performance on structured tasks. These reveal how hardware error rates and architecture impact actual algorithm steps: Small physical gate errors can have outsized effects in structured circuits (like QFTs), compared to random gates~\cite{calderon2023quantum}. This insight goes beyond what isolated gate fidelity tells us by showing how errors compound in a realistic workload. Meanwhile, mid-level benchmarks also highlight the effectiveness of software techniques (compilation, error mitigation, error correction) in harnessing the hardware. Even if hardware error rates are high, clever mid-level strategies might extend the effective circuit depth. Conversely, if the overhead of error correction is too large, it can erase any theoretical speedup an algorithm had. Thus, the mid-level mediates how hardware qualities translate into algorithm performance, and improvements at this level (better compilers, error mitigation) directly raise the ceiling set by raw hardware capabilities.

High-level application benchmarks, in turn, integrate and stress-test the entire stack, often exposing interplay that low-level tests might miss. Because these benchmarks involve running a full problem (albeit at small scale), they verify whether the gains in hardware and software actually deliver an advantage on real tasks. A ground-state energy VQE benchmark might succeed on a simple molecule, but as the molecule’s size grows, it could fail to converge due to accumulated gate errors or insufficient circuit depth – reflecting the limits indicated by mid-level volumetric metrics. Notably, application-driven metrics also provide feedback to lower levels: If the individual metrics with a focus on lower levels in the hardware stack are very good but the computer fails with regard to the application-based benchmarks, this either indicates incorrect benchmarks in the lower part of the stack or weak points that have not yet been benchmarked, especially in the interaction of the components. In this way, application-level benchmarks inform hardware requirements, creating a design feedback loop. As one study notes, algorithmic performance results can guide the development of quantum hardware by linking observed algorithm outcomes to needed hardware advances~\cite{georgopoulos2021quantumcomputerbenchmarkingquantum}, making a step in the direction of ``codesign'' approaches. Similarly, these benchmarks emphasize system-wide issues like orchestration, calibration drifts, or multi-qubit crosstalk that might be averaged out or overlooked in simpler tests. In summary, high-level benchmarks validate the efficacy of lower-level improvements in a real-world context and highlight any remaining gaps.

In the following, we would like to highlight a few dependencies in particular:

\begin{itemize}
    \item Gate fidelity → Algorithm success: Improvements in low-level gate fidelity (e.g. measured via RB or tomography) directly increase the depth and complexity of circuits that can run before failure. Conversely, high error rates set a hard limit on achievable algorithm accuracy and runtime, as seen by success probability decays in both random sequences and structured algorithm tests.
    \item Hardware architecture constraints → Software overhead: Physical constraints like qubit count and connectivity dictate compilation overhead at the mid-level (e.g. additional SWAP gates for limited topologies), which in turn raises error rates in applications. A benchmark of an algorithm on different hardware may show considerably better performance when topology-aligned, underscoring this cross-layer dependency.
    \item Mid-level optimization → Effective performance: The efficacy of mid-level techniques (error correction, circuit optimization) moderates the translation of hardware metrics into application outcomes. Achieving a practical quantum advantage may hinge on reducing error-correction overhead or improving compiler efficiency.
    \item Structured vs. random circuits: Benchmarks also reveal that error impact differs by workload. Randomized benchmarks give an average hardware error rate that is agnostic to specific algorithms, whereas algorithm-specific benchmarks may show worst-case sensitivities~\cite{hashim2023benchmarking}. This pattern warns that solely relying on one type of benchmark can be misleading; a holistic view requires both randomized and structured tests.
    \item End-to-end validation and feedback: Finally, application-level benchmarks tie everything together, confirming whether lower-level improvements yield system-level benefits. However, it must also be noted here that individual application-based benchmarks can only benchmark the performance of the underlying hardware for the applied problem class and that these results cannot necessarily be transferred to other problem classes.
\end{itemize}

The interconnectedness observed implies that no single benchmark level suffices – effective quantum evaluation demands a stack-wide perspective. Low-level benchmarks provide essential component metrics, mid-level benchmarks translate those into circuit-level performance, and high-level benchmarks assess the ultimate utility of the system. Together, they form a complementary set: Each addresses limitations of the others. A key conclusion is that progress in QC will require coordinated improvements across all levels and layers of this hierarchy, as well as cross-talk between benchmarking methods. Benchmark developers are increasingly aware of this synergy, as seen in integrated frameworks that mirror the quantum stack. Thus, the three benchmarking categories we identified should not be viewed in isolation; rather, they form a complementary set, each addressing limitations of the others, and all are needed for a comprehensive evaluation.

\section{Conclusion}\label{conclusion}
In this review, we presented a structured taxonomy with definitions of QC benchmarking techniques, organized across the hardware, software, and application levels. We explicitly align our classification with stakeholder needs, reflecting the diverse objectives across the QC stack. Through this framework, we demonstrated that no single benchmark can comprehensively assess all aspects of QC performance -- highlighting the necessity for multiple, purpose-specific benchmarking methods. Moreover, by analyzing the interdependencies between benchmarking approaches, we showed that many methods serve complementary roles, where advancements at one level must be interpreted in the context of others. This holistic perspective is essential for developing fair, informative, and future-ready quantum benchmarks.

Beyond its descriptive contributions, this review provides a conceptual foundation for improving communication and collaboration in the QC community. By linking benchmark categories to stakeholder groups, it enables developers to tailor metrics to actual needs while promoting a shared language across disciplines. The analysis of interdependencies encourages a shift from isolated assessments toward integrated, system evaluation, fostering ``a stack-wide mindset''. Progress is urged to be validated across the full stack —from low-level hardware metrics to high-level application outcomes— highlighting the importance of cross-layer effects in the stack, which remain underrepresented in current benchmarking practice. As a result, claims of advancement will be more rigorously scrutinized at multiple levels before being accepted.

Despite all efforts, several limitations must be acknowledged. First, our focus was restricted to universal, gate-based QC, excluding alternative paradigms such as quantum annealing or analog quantum simulation. Consequently, the proposed taxonomy may not generalize to those models. Second, given the rapid pace of development in the field, some recent or non-peer-reviewed benchmarks may not have been captured, despite our forward-backward search strategy. Finally, although the taxonomy was built using NLP-based clustering refined through manual evaluation, the categorization and interpretation of borderline cases inherently involve a degree of subjectivity.

Looking ahead, several research directions emerge from this work. First, future efforts should extend the taxonomy to encompass other QC paradigms —such as quantum annealers and analog quantum simulators— to determine whether new benchmark categories or principles are required, or whether the existing framework can be meaningfully adapted. Research should also explore the development of composite, cross-level benchmarks that explicitly link low-level metrics to high-level application performance, enabling more accurate assessments of system-wide progress and guiding development priorities. This approach would also deepen the understanding of the interconnectedness across the stack and help identify bottlenecks and propagation effects. Lastly, given the rapid pace of innovation, a continuously updated repository or living survey would ensure that benchmarking methodologies remain up-to-date, comprehensive, and aligned with emerging technologies.

Even if the comparison and benchmarking of QC systems and their components sound simplistic and superficial at first, a fair comparison, taking into account various circumstances, is a fine art that requires a great deal of knowledge and experience. Benchmarks should not only be considered in isolation for themselves but have a strong interdependency on the development of new algorithms, components, and hardware. Without fair comparisons, neither weaknesses nor progress can be tracked in the field, leaving the direction of research work undefined. With our paper on QC benchmarking, we have made a contribution to a better understanding of the topic, which will help other researchers to better assess which benchmark is well suited for their purpose. 

\newpage
\section*{List of Abbreviations}
\begin{table}[h!]
\centering
\begin{tabular}{ll}
\textbf{Abbreviation} & \textbf{Definition} \\
\hline
ABBs & Algorithm-based Benchmarks \\
AGI & Average Gate Infidelity \\
AQC & Adiabatic Quantum Computing \\
BAS & Bars and Stripes \\
CB & Cycle Benchmarking \\
CS-QPT & Compressed Sensing QPT \\
CS-QST & Compressed Sensing QST \\
DDQCL &  Data-driven Quantum Circuit Learning \\
DFE & Direct Fidelity Estimation \\
DM & Density Matrix \\
Direct RB & Direct Randomized Benchmarking \\
DRB & Dihedral Randomized Benchmarking \\
FCS & Finitely Correlated States \\
GD-QPT & Gradient Descent QPT \\
GD-QST & Gradient Descent QST \\
GST & Gate Set Tomography \\
HDBSCAN & Hierarchical Density-Based Spatial Clustering of Applications with Noise \\
IRB & Interleaved Randomized Benchmarking \\
MaxCut & Maximum Cut \\
MLE & Maximum Likelihood Estimation \\
MPS & Matrix Product States \\
NISQ & Noisy Intermediate-Scale Quantum \\
NLP & Natural Language Processing \\
NN & Neural Network \\
PI & Permutationally Invariant \\
PLSE & Projected Least Squares Estimator \\
QAOA & Quantum Approximate Optimization Algorithm \\
QBM & Quantum Benchmarking \\
QC & Quantum Computing \\
QCVV & Quantum Characterization, Verification, and Validation \\
QEC & Quantum Error Correction \\
QFT & Quantum Fourier Transform \\
QML & Quantum Machine Learning \\
QPT & Quantum Process Tomography \\
QPU & Quantum Processing Unit \\
QST & Quantum State Tomography \\
QUBO & Quadratic Unconstrained Binary Optimization \\
RB & Randomized Benchmarking \\
RBM & Restricted Boltzmann Machine \\
RBM-QST & Restricted Boltzmann Machine QST \\
SDKs & Software Development Kits \\
SPAM & State Preparation and Measurement \\
ST & Shadow Tomography \\
UMAP & Uniform Manifold Approximation and Projection \\
VQE & Variational Quantum Eigensolver \\
XEB & Cross Entropy Benchmarking \\
\hline
\end{tabular}
\caption{List of abbreviations used in this paper.}
\label{tab:abbreviations}
\end{table}

\newpage

\section{Declarations}

\subsection{Competing interests}
The authors declare that they have no competing interests.

\subsection{Acknowledgements}
This paper was partially funded by the Federal Ministry of Research, Technology and Space through the funding program “quantum technologies – from basic research to market” (contract number: 13N16196). J.S. and S.E. acknowledges support from the German Federal Ministry for Economic Affairs and Climate Action through the funding program “Quantum Computing – Applications for the industry” based on the allowance “Development of digital technologies” (contract number: 01MQ22008A). F.H.R. acknowledges support by the DAAD programme Konrad Zuse Schools of Excellence in Artificial Intelligence, sponsored by the Federal Ministry of Research, Technology and Space.

\newpage

\begin{appendices}

\section{BERTopic Initial Clustering}\label{BERTopic}
After collecting the initial corpus of literature for the present work, we utilized an initial clustering method to construct preliminary benchmark categories. This method was based on the technique of topic modelling, in particular through the transformer-based \texttt{BERTopic} library~\cite{grootendorst2022bertopic}.

\begin{figure}
    \centering
    \includegraphics[width=0.7\linewidth]{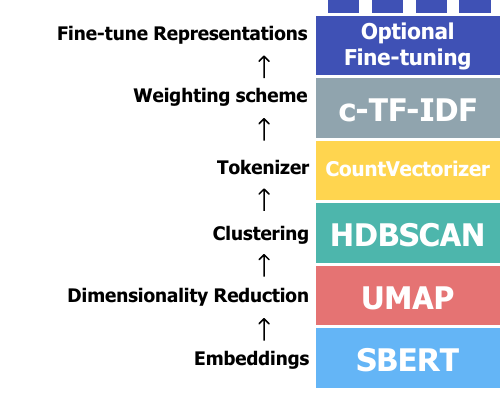}
    \caption[BERTopic Overview]{Overview of the different components of the topic modelling process through BERTopic. The bottom-up process on the left shows the order of execution for the 6 main abstract steps of the process. The bricks on the right show concrete instances or algorithms that are most commonly used for each of the abstract processes, which differ slightly from our implementation, where the Embeddings were provided by the \texttt{all-MiniLM-L6-v2} model, and a fine-tuning of labels was provided by the \texttt{gpt-4-0125-preview} model. The brick representation is used to indicate that each concrete process can be replaced for an alternative and stacked freely.}
    \label{fig:bertopic-algorithm}
\end{figure}

\definecolor{orange0}{RGB}{245, 194, 142}
\definecolor{green1}{RGB}{161, 205, 154}
\definecolor{red2}{RGB}{222, 151, 149}
\definecolor{purple3}{RGB}{197, 180, 219}
\definecolor{brown4}{RGB}{193, 171, 166}
\definecolor{pink5}{RGB}{233, 189, 222}
\definecolor{grey6}{RGB}{191, 191, 191}
\definecolor{lgreen7}{RGB}{221, 222, 153}
\definecolor{cyan8}{RGB}{158, 220, 229}
\definecolor{blue9}{RGB}{152, 186, 222}

\begin{figure*}
    \centering
    \includegraphics[width=0.95\linewidth]{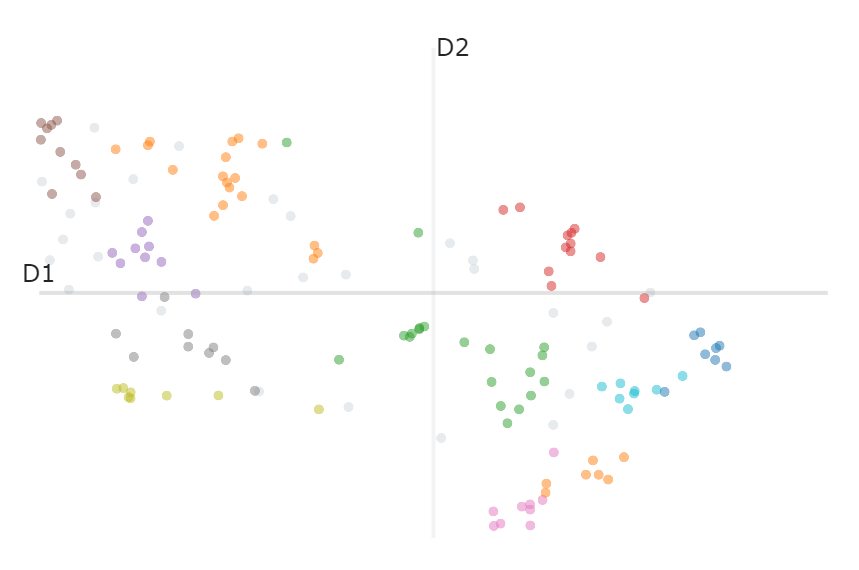}

    \begin{tabular}{| m{0.06\linewidth} | m{0.87\linewidth} |}
    \hline
    Color & Cluster Label \\\hline
    \cellcolor{orange0}0 & Quantum Circuit Simulation and Sampling \\
    \cellcolor{green1}1  & Characterizing Quantum Computers through Randomized Benchmarking and Noise Analysis \\
    \cellcolor{red2}2  & Quantum Process Tomography \\
    \cellcolor{purple3}3  & QPACK Benchmarking for Quantum Computers \\
    \cellcolor{brown4}4 & Benchmarking and Performance Analysis of Quantum Algorithms and Solvers \\
    \cellcolor{pink5}5 & Quantum Entanglement and Superconducting Qubits \\
    \cellcolor{grey6}6  & Benchmarking and Optimization of Quantum Circuits and Algorithms \\
    \cellcolor{lgreen7}7  & Machine Learning and Quantum Datasets \\
    \cellcolor{cyan8}8  & Quantum Gate Fidelity and Benchmarking \\
    \cellcolor{blue9}9  & Characterization and Benchmarking of Noisy Quantum Circuits and Qubits \\
    \cellcolor{orange0}10  & Quantum Computing with Silicon Spin Qubits \\
    \cellcolor{green1}11  & Quantum Chemistry and Molecular Basis \\\hline
    \end{tabular}

    \caption[BERTopic Clustering Results]{
    Results of the BERTopic clustering of the collected literature. The visualization is a 2D projection of the higher-dimensional elements of each cluster. Each color represents a different benchmark category. As shown in the table above, both the clusters 0 and 10 and 1 and 11 were assigned the same color. Cluster 0, labelled as \textit{Quantum Circuit Simulation and Sampling}, corresponds to the orange points on the upper-left quadrant. The green points close to this cluster correspond to the label \textit{Quantum Chemistry and Molecular Basis}, as well as the very dense green points near the center of the figure. The orange cluster on the bottom left of the image corresponds to label 10, \textit{Quantum Computing with Silicon Spin Qubits}, close to cluster 5, \textit{Quantum Entanglement and Superconducting Qubits}. Cluster 1, corresponding to the label \textit{Characterizing Quantum Computers through Randomized Benchmarking and Noise Analysis}, corresponds to the green points on the lower right quadrant.
    }
    \label{fig:bertopic-results}
\end{figure*}

Before performing the topic modelling analysis, we preprocessed the articles to optimize the quality of the generated clusters. First, we removed the most common words that were not relevant for the identification of distinct benchmark categories and could have led to the creation of clusters based on other properties:

\begin{quote}
    \textit{Journal, author, university, department, departments, volume, year, letters, articles, references, phys, rev, vol, et, al, physical review, latexit, arxiv.}
\end{quote}

In addition to these words, we removed the English stop words contained in the \texttt{NLTK} library~\cite{bird2009nltk}, highly common and generic words in English literature, such as demonstratives and personal pronouns. To complete the preprocessing step, we removed all digits from the corpus and finally tokenized the filtered corpus using the word tokenizer provided by \texttt{NLTK}.

For the clustering process, we created numerical representations of the filtered corpus. The sentence transformer model \texttt{all-MiniLM-L6-v2}, which is a fine-tuned version\footnote{More details about the fine-tuning available at: https://huggingface.co/sentence-transformers/all-MiniLM-L6-v2} of Ref.~\cite{10.5555/3495724.3496209}, was used to embed each of the articles into a numerical representation, as this excels at capturing semantic similarity between documents. The model was accessed through the \texttt{Sentence-Transformers} library \cite{reimers-2019-sentence-bert}. The embeddings produced by this model have a dimensionality of $384$ features, creating a compact representation compressing the meaning of each document. The high dimensionality can be challenging for clustering methods. To address this issue, the \textit{Uniform Manifold Approximation and Projection} (UMAP) dimensionality reduction technique~\cite{McInnes2018} was used. Finally, the \textit{Hierarchical Density-Based Spatial Clustering of Applications with Noise} (HDBSCAN) algorithm \cite{McInnes2017} was used to create the initial clusters. As the name suggests, this algorithm constructs a hierarchy of clusters by analysing the density of data points in the embedding space, and then condenses this hierarchy to extract the most stable clusters, identifying clusters of different shapes and densities while labelling outliers as noise

To create more interpretable topic representations corresponding to the generated clusters, the \texttt{CountVectorizer} class provided by the \texttt{scikit-learn} library \cite{scikit-learn} was used. This vectorizer creates n-grams of length two from the clusters, capturing common word pairs that characterize each topic. Additionally, English stop words were removed from these labels. To create even clearer and more interpretable labels for the topics, the \texttt{gpt-4-0125-preview} \cite{openai2024gpt40125} model was used through the OpenAI API, taking both the initial cluster labels and the corresponding articles as inputs, and producing more refined labels as outputs. An overview of the topic modelling process through BERTopic and the final results can be seen in Fig.~\ref{fig:bertopic-algorithm} and Fig.~\ref{fig:bertopic-results} respectively.

\end{appendices}

\end{document}